\documentclass[twocolumn]{aastex63}

\usepackage[utf8]{inputenc}
\usepackage{hyperref}
\usepackage{silence}
\usepackage{graphicx}
\usepackage[caption=false]{subfig}

\usepackage{float}
\usepackage{color,soul}

\usepackage{wrapfig}
\usepackage{bm}
\usepackage{amsmath}

\usepackage{fancyhdr}
\usepackage{wasysym}
\usepackage{natbib}

\begin{document}

\title{Robustness of Synthetic Observations in Producing Observed Core Properties: Predictions for the TolTEC Clouds to Cores Legacy Survey}

\author[0000-0002-8667-6428]{S. K. Betti}
\affiliation{Department of Astronomy, University of Massachusetts, Amherst, MA 01003, USA}

\author[0000-0002-6447-899X]{R. Gutermuth}
\affiliation{Department of Astronomy, University of Massachusetts, Amherst, MA 01003, USA}

\author[0000-0003-1252-9916]{S. Offner}
\affiliation{Department of Astronomy, University of Texas at Austin, Austin, TX 78712, USA}

\author{G. Wilson}
\affiliation{Department of Astronomy, University of Massachusetts, Amherst, MA 01003, USA}

\author{A. Sokol}
\affiliation{Department of Astronomy, University of Massachusetts, Amherst, MA 01003, USA}

\author[0000-0002-0557-7349]{R. Pokhrel}
\affiliation{Department of Astronomy, University of Toledo, Toledo, OH 43606, USA}

\correspondingauthor{S. K. Betti}
\email{sbetti@umass.edu}

\begin{abstract}
We use hydrodynamical simulations of star-forming gas with stellar feedback and sink particles (proxies for young stellar objects, i.e., YSOs) to produce and analyze synthetic 1.1~mm continuum observations at different distances (150 - 1000~pc) and ages (0.49 - 1.27 Myr). 
We characterize how the inferred core properties, including mass, size, and clustering with respect to diffuse natal gas structure, change with distance, cloud evolution, and the presence of YSOs.  We find that atmospheric filtering and core segmentation treatments have distance-dependent impacts on the resulting core properties for $d <$~300~pc and 500~pc, respectively, which dominate over evolutionary differences.  Concentrating on synthetic observations at further distances (650-1000~pc), we find a growing separation between the inferred sizes and masses of cores with and without YSOs in the simulations, which is not seen in recent observations of the Mon~R2 cloud at 860~pc.  We find that the synthetic cores cluster in smaller groups, and their mass densities are correlated with gas column density over a much narrower range, than the MonR2 observations. Such differences limit applicability of the evolutionary predictions we report here and motivate future efforts to adapt our synthetic observation and analysis framework to next generation simulations such as STARFORGE. These predictions and systematic characterizations will help guide analysis of cores for the upcoming TolTEC Clouds to Cores Legacy Survey on the Large Millimeter Telescope Alfonso Serrano (LMT). 
\end{abstract}


%-------------------------------------------------------------------------------------------------------------------------

\section{Introduction}

Dense cores within giant molecular clouds (GMC) are the birthplaces of stars \citep{Bergin2007, diFrancesco2007, Ward-Thompson2007}.  These cores are part of a hierarchical structure of fragments within GMCs \citep{Pokhrel2018} that form as a result of a variety of physical processes, including self-gravity \citep{Heyer2009, Ballesteros-Paredes2011, Ballesteros-Paredes2012},  magnetohydrodynamic turbulence \citep{MacLow2004, Hennebelle2012}, supersonic interstellar turbulence \citep[e.g.][]{Pudritz2013}, and ionization of molecular gas \citep{Whitworth1994, Dale2009}.  The densest filamentary structures within clouds have been observed to host cores \citep{Andre2010, Polychroni2013} of order 10$^{4-5}$ cm$^{-3}$ in density, 0.03-0.1 pc in size, and T~$<$~12 K in temperature \citep{diFrancesco2007}.  Through gravitational instabilities, some prestellar cores may collapse to form protostars.  

In the millimeter to submillimeter regime, thermal emission from cores is optically thin, which allows the flux density to trace the total dust mass, making them ideal targets of study for both ground-based and space-based submm telescopes. 
Various surveys of nearby Gould Belt molecular clouds (d $<$ 500 pc) have produced censuses of dense cores \citep[e.g.,][]{Konyves2015, Marsh2016, Bresnahan2018, Benedettini2018, Ladjelate2020, Enoch2006, Enoch2007, Enoch2008}.  These surveys also confirmed that the cores predominantly are found within relatively dense $\sim0.1$~pc filamentary structures in clouds \citep{Arzoumanian2011, Arzoumanian2019, Andre2010, Andre2014, Menshchikov2010,Polychroni2013}.  These cores have been observed in varying stages of gravitational stability, from those that are gravitationally collapsing and forming YSOs to those that are pressure-confined but at too low masses to collapse further under self-gravity \citep{Kirk2016, Kirk2017, Friesen2017}.

Overall, the mass distribution of cores (core mass function, CMF) has been shown to follow the same shape as the stellar initial mass function (IMF) shifted to 1-5$\times$ in mass in a variety of clouds \citep{Andre2010, Konyves2015, Marsh2016, Bresnahan2018, Benedettini2018, Konyves2019, Sokol2019, Ladjelate2020}.  This similarity has been interpreted as evidence that the CMF sets the distribution of stellar masses \citep{Offner2014b}. However, most core surveys remain incomplete at masses $<$ 1 M$_\odot$, so it is difficult to determine whether the CMF turnover is similar to that of the IMF and likewise invariant with environment \citep{Offner2014b, Guszejnov2015}.
The incomplete sampling of the CMF at the low mass end is a result of limited sensitivity and angular resolution of previous measurements, limiting their ability to distinguish dense cores from larger natal gas structures, like filaments.  Along with sensitivity, the various Gould Belt surveys only observed clouds within 500 pc, limiting their view to only a narrow range of star-forming environments.  Surveys of more distant and active star-forming clouds, including Cygnus-X \citep{Cao2019} and Mon~R2 \citep{Sokol2019}, give insights into more active star formation but also suffer from low sensitivity and angular resolution.    
    
In order to address these issues, the TolTEC Clouds to Cores Legacy Survey (C2C) will survey cores in 10 clouds at a variety of ages and distances with the new TolTEC three band submillimeter (submm) imaging polarimeter on the Large Millimeter Alfonso Serrano Telescope \citep[LMT;][]{Wilson2020}.  Under optimal performance characteristics for the LMT and TolTEC, C2C plans to survey 88 deg$^2$ of nearby molecular clouds in 100 hours of observations, with the goal of reaching 0.24 mJy / beam RMS at 1.1 mm in order to detect cores with 0.1 M$_\odot$ at 4$\sigma$.  
The survey will observe thousands of spatially resolved cores to a uniform surface brightness, and therefore mass, from Ophiuchus (at 137 pc) to Cygnus-X (1400 pc) as a result of the high angular resolution (5$\arcsec$ at 1.1 mm) and fast mapping speed (2-12 deg$^2$ / mJy$^2$ / hr).  This will allow for a complete characterization of the CMF down to $<0.3$ M$_\odot$ and will place potent new constraints on the origin of the CMF, how it potentially varies with natal environmental conditions, and how or if it influences the stellar IMF.

Numerical simulations of core formation play a key role for interpreting observations, including understanding the potential relationship between the CMF and IMF and the role of environmental processes in shaping cores \citep{Lee2020}.  Previous works have predicted a range of relationships between the IMF and CMF, from no link between the two \citep{Bonnell2001, Bate2003, Clark2007}, to the IMF directly mapping from the core mass distribution \citep{Padoan2002, Hennebelle2008, Hennebelle2009, Oey2011, Hopkins2012, Guszejnov2015}.  
Simulations have shown that the CMF and resulting IMF are sensitive to a variety of physical and environmental processes.
These include magnetic fields limiting fragmentation \citep{Padoan2007, Commercon2011, Hennebelle2011, Myers2013}, radiative feedback heating the surrounding material and thereby increasing the local Jeans mass \citep{Bate2009,Offner2009,Krumholz2011,Krumholz2016}, kinematic feedback such as outflows and winds, which reduce the star-formation efficiency of dense cores \citep{Krumholz2012,Cunningham2018,Guszejnov2020}, and turbulence that decreases the local free-fall time and preventing collapse \citep{Robertson2012, Murray2017}.  When comparing their results to observations, most of these simulations do not take into account observational biases, such as atmospheric filtering, noise, and image segmentation algorithms which affect observational results.  They also focus on snapshots in time and rely on sink particles or an overdensity mass to define a core.  Recent work by \citet{Smullen2020} investigated the evolution of cores over time in magnetohydrodynamic (MHD) simulations in order to explore how gas reservoirs evolve and how core identification by a commonly-used structure finding algorithm based on dendrograms may  bias core properties. They found that, while the distributions of core properties and the CMF remained relatively stable in time, the boundaries of individual cores fluctuated significantly.   
Although a variety of simulations have explored environmental effects on the formation of cores 
there has been less attention devoted to using synthetic observations as a tool for a) testing the robustness of simulations in reproducing observational core properties, along with systematics in observational and analytic reduction tools (data reduction pipelines and core identification analysis), and b) determining how the above environmental effects influence the observed core properties.  

Due to the extensive nature of the C2C survey, the properties of cores surveyed will be influenced by their environment and evolutionary status, as stellar feedback, magnetic field, and gas accretion can all play a role in core formation and evolution \citep{Kirk2013, Pokhrel2018}.  In order to compare cores at varying distances and ages, and interpret the core properties, including mass, size, clustering, and star-forming yield, predictions of core properties are necessary to distinguish the underlying mechanisms.  However, without a detailed analysis and characterization of the robustness of both simulations and analysis tools we will not be able to use synthetic observations or simulations to accurately interpret how core formation varies in different environments in the various molecular clouds in the C2C survey.  

In this paper, we test the reduction and analysis tools that are currently planned for use on C2C for their ability to identify and characterize cores uniformly at various distances and ages.  We will also probe the robustness of synthetic 1.1~mm emission maps of star-forming molecular gas regions from numerical radiative transfer simulations in reproducing observed core properties. In Section \ref{sec2}, we describe the hydrodynamical radiative transfer simulations, the observational data, and the construction of the synthetic emission maps.  In Section \ref{sec3}, we describe the core identification and property characterization algorithms.  In Section \ref{sec4}, we assess the core properties and systematics found from the synthetic observations.  In Section \ref{sec5}, we compare synthetic observations to observations of the Mon~R2 GMC in order to test the robustness of the simulation in reproducing observed core properties.  In Section \ref{sec6}, we explore the evolution of observable core properties (mass, size, clustering) by extracting and characterizing cores in three snapshots in time from one simulation.  We conclude by summarizing our results in Section \ref{sec7}.    

\begin{deluxetable*}{ccccccccc}[tb]
\tablewidth{0pt}  
\tablecaption{Simulation Properties \label{sim_prop}
}
\tablehead{\colhead{Model$\tablenotemark{a}$} & \colhead{$M$} & \colhead{$\mathcal{M}$} & \colhead{$L$} & \colhead{Age} & \colhead{$T_{i}$}  & \colhead{$N^3$} & \colhead{$\Delta$x$_{min}$}  & \colhead{$N_{sink}$} \\ \colhead{} & \colhead{($ 10^4 M_\odot$)} &  &\colhead{(pc)} & \colhead{(Myr)} & \colhead{(K)}  & \colhead{} & \colhead{(pc)}  & \colhead{}}
\startdata
RT1   &   1.425   &  14 &  10   &   0.83    &   10  &   512$^3$ &   0.005    & 169   \\
RT2.1 &   0.378   & 10.5 &  5    &   0.49    &   10  &   256$^3$ &   0.001    & 3 \\
RT2.2 &   0.364   & 10.5&  5    &   0.92    &   10  &   256$^3$ &   0.001    & 62   \\
RT2.3 &   0.327   & 10.5& 5    &   1.27    &   10  &   256$^3$ &   0.001    & 120   \\
\enddata
\tablenotetext{a}{Model name, gas mass, Mach number, domain length, analysis output time, initial gas temperature, base-grid size, minimum cell size, and number of sink particles.}
\end{deluxetable*}

\section{Simulations, Observations, and Their Synthesis} \label{sec2}

Our goal is to produce synthetic mm-wave continuum observations from projected views of simulations of star-forming molecular gas for core extraction and analysis as if they were observed in the real sky. Here we describe the simulations and the LMT and AzTEC 1.1~mm observations used to make our synthetic observations, followed by the process that we follow to make them.

\subsection{Radiative Transfer Simulations}

We use hydrodynamic simulations produced using \texttt{Orion}, an adaptive mesh refinement (AMR) code that follows the equations of hydrodynamics, including self-gravity and grey flux-limited diffusion radiative transfer \citep[RT; see][for full details of the numerical methods employed]{Klein1999,Krumholz2004,Krumholz2007}.  These simulations include star particles that follow a sub-grid prescription of protostellar evolution, including radiative feedback due to accretion and nuclear processes \citep{Offner2009} and are meant to simulate low-mass star formation in a turbulent molecular cloud.  

We analyze two RT simulations with different domain sizes and resolutions (see Table \ref{sim_prop}). Both simulations adopt periodic boundary conditions. The larger simulation (RT1) has a domain size of 10 pc, Mach number of 14, a base grid of 512$^3$, 2 AMR levels of refinement, and a maximum resolution of 1030 AU (previously also analyzed in \citealt{Qian2015}).  The smaller simulation (RT2) has a domain size of 5 pc, Mach number of 10.5, a base grid of 256$^3$, 4 AMR levels of refinement, and a maximum resolution of 126 AU.  Both simulations have an initial gas temperature of 10 K.  

The simulations are initialized by applying velocity perturbations of the form $P(k) \propto k^0$ to an initially uniform density field \citep{MacLow1999} with input normalized wavenumbers in the range $k\sim1-2$. 
The perturbations are continued until a turbulent steady state is reached and the gas distribution  follows $P(k) \propto k^{-2}$ (at approximately two crossing times) at which time self-gravity is turned on within the simulation.  Energy is continually injected to ensure the turbulence remains steady and does not decay as a result of shock dissipation.     
At each time-step of the simulation, the radiative transfer equation in a gray flux-limited diffusion approximation is solved to determine the radiation energy density \citep{Krumholz2007}.  The gravitational potential is also calculated from the Poisson equation \citep{Klein1999}, and the radiation, including radiation feedback from sink particles (forming stars) is updated \citep{Offner2009}.  

Once self-gravity is turned on, collapse can commence. New AMR grids are added when the density in a collapsing region exceeds the critical Jeans density for a Jeans number of N$_\mathrm{J}>$  0.25  \citep{Truelove1997}. If the critical Jeans density is exceeded on the maximum AMR level, a sink particle is inserted \citep{Krumholz2004}. 
When this occurs, the excess mass within the dense cells is transferred to the inserted sink particle.  The sink particles interact with the gas through mass accretion within the nearest four cells and through gravitational attraction; however, all knowledge of the hydrodynamics within the cell is lost, and the gas will continue to collapse. In these calculations, each sink particle represents an individual star system; only wide binaries, i.e., those with initial separations $d\gtrsim 800$ AU, are resolved.

\subsection{Observational Data}

We use 1.1~mm continuum data from \citet{Sokol2019} for two purposes in this work.  First, they are the foundation of the synthetic observations, providing actual realizations of atmospheric emission and other observational systematics once the Mon~R2 emission is removed (described below). Second, the core census of Mon~R2 provides an important comparison set for the analyzed synthetic observations.  The original observations were taken between 2014 November 27 and 2015 January 31 with AzTEC, a 144 element 1.1 mm bolometer array \citep{Wilson2008, Austermann2009} during the 32~m diameter early science configuration on the 50~m diameter LMT. Fourteen fields of Mon~R2 were mapped covering a region 2~deg$^2$.  Noise levels of $\sim7$~mJy per beam rms are determined from the \citet{Scott2008} jackknifing technique (see \citet{Sokol2019} for a coverage map of the fields and full description of the observations).  The maps (both the original observations and the synthetic observations) were reduced using the standard AzTEC C++ reduction pipeline described in \citet{Sokol2019} \citep[\texttt{macana;}][]{Scott2008}.  The final maps have an angular resolution of $\sim$12" FWHM, corresponding to 0.05 pc at the distance of Mon~R2 \citep[860 pc; derived from][]{GAIA2018, Pokhrel2020}.

\subsection{Synthetic Observations} 

We use one snapshot from RT1, and three snapshots from RT2 (see Table \ref{sim_prop}) to produce the synthetic observations.  These snapshots correspond to different output times from the simulations once gravity is turned on ($t=0$).  At the output time of RT1, 0.83~Myr, there are 169 sink particles.  For the three RT2 snapshots, RT2.1 (0.49~Myr), RT2.2 (0.92~Myr), and RT2.3 (1.27~Myr), there are 3, 62, and 120 sink particles, respectively.  The sink particles are proxies for YSOs. Their locations indicate where the density has exceeded the Jeans number and collapse occurred until eventual star formation, i.e., they mark the locations of protostellar cores.   

We convert the simulation gas to 1.1~mm continuum synthetic emission maps at the resolution of the Mon~R2 AzTEC maps (1\arcsec\ per pixel) and a representative range of distances for the C2C clouds (150-1000~pc), while preserving the flux and physical scale of the simulations. \

We first flatten the snapshot gas density cubes onto a fixed grid (RT1: 2048$\times$2048 pixels, RT2.1-2.3: 4096$\times$4096 pixels).  The projected density of the resulting 2D map is given a world coordinate system where the center pixel of the map corresponds to the center pixel from one field of the AzTEC data.  
We use Field~4 centered on RA (J2000): 06$^{\rm h}$07$^{\rm m}$59$^{\rm s}$.89 and Dec (J2000): -07$^{\circ}$00${'}$04${''}$.1 to map our simulation.  
This field is large (40\arcmin~$\times$~40\arcmin) with a noise level of 7~mJy/beam and a maximum peak flux of 109~mJy/beam, making it an ideal candidate to insert synthetic emission maps.  The density maps are scaled to the resolution necessary for the synthetic observation to appear at a given distance in the sky.  
The simulations are then reprojected onto the same grid as Field~4 while preserving the flux and scale.  The 3D Cartesian sink particle positions are similarly flattened and projected onto the same world coordinate grid.  

As noted above, the simulations are 5 and 10 pc across, and Field~4 is larger than that at the Mon~R2 distance (or further). To address this, we exploit the periodic boundary conditions of the simulations and tile the projected density map in order to fill the whole field.  We next apply a Gaussian kernel the same size as the LMT (32~m) and AzTEC natural beam (FWHM$\sim8.5''$) to smooth the projected maps to the resolution of the AzTEC maps.  

To determine the thermal dust emission at 1.1 mm (in Jy/beam), we assume the dust and gas are thermally coupled \citep{Offner2009}, and the gas density is optically thin ($\tau \sim$ 0.005 for RT1 and RT2).  Therefore, we can apply the radiative transfer equation for an optically thin gas 
to convert from projected gas density (g cm$^{-2}$) to dust emission (Jy/beam):
\begin{equation}
I_\nu = B_\nu (T_d) \tau_\nu, 
\end{equation} 
where $I_\nu$ is the emission, $B_\nu(T_d)$ is the Planck function for a dust temperature T$_d$, and $\tau_\nu$ is the optical depth of the simulated cloud.  The dust temperature, T$_d$, is the temperature at each map pixel found from the radiation energy of the projected simulation.   
Since $\tau_\nu = \kappa_\nu \Sigma$, where $\Sigma = M / D^2$ is the mass surface density and $\kappa_\nu$ is the dust opacity, then the emission at 1.1 mm is
\begin{equation}  
I_\mathrm{1.1\ mm} = B_\mathrm{1.1\ mm} (T_d) \kappa_\mathrm{1.1\ mm} \left(\frac{M }{D^{2}}\right).
\end{equation}
We assume a constant dust opacity of 0.0121 cm$^{2}$ g$^{-1}$ at 1.1 mm \citep[model 4 opacities;][]{Ossenkopf1994} that includes a gas-to-dust ratio of 100.  $M$ is the gas mass and $D$ is the distance corresponding to each model distance.  This converts the gas density from g cm$^{-2}$ to Jy cm$^{-2}$.  In order to convert to Jy/beam we multiply by the beam in cm$^{2}$/beam.

\begin{deluxetable*}{cccc||cccc}[t]
\tablewidth{0pt}  
\tablecaption{\texttt{macana} Runs $\&$ ImSeg Results \label{modelruns}
}
\tablehead{\colhead{Model} &\colhead{distance} & \colhead{view} & \colhead{resolution$\tablenotemark{a}$} & \colhead{N$_{\rm cores}$} & \colhead{N$_{\rm starred\ cores}$} & \colhead{f$_{sink}\tablenotemark{b}$} & \colhead{f$_{starred}\tablenotemark{c}$} \\ {} & {(pc)} & {} & {($''$)} & {} & {}  & {} & {}}
\startdata
\hline 
RT1\_D860z & 860 & $z$ & 1.21   &  222 & 78 & 0.53 & 0.35  \\
\hline
RT2.1\_D150z & 150 & $z$ & 1.68 & 146 (146) & 1 (1) & 0.50 (0.50) & 0.01 (0.01)   \\
RT2.1\_D300z & 300 & $z$ & 0.84 & 83 (165) & 2 (2) & 1.00 (1.00) & 0.01 (0.01)   \\
RT2.1\_D450z & 450 & $z$ & 0.56 & 41 (206) & 2 (2) & 1.00 (0.50) & 0.01 (0.05)   \\
RT2.1\_D650z & 650 & $z$ & 0.38 & 21 (198)  & 2 (4) & 1.00 (0.80) & 0.09 (0.02)   \\
RT2.1\_D860z & 860 & $z$ & 0.30 & 11 (178)  & 2 (5) & 1.00 (0.38) & 0.18 (0.03)  \\
RT2.1\_D1000z & 1000 & $z$ & 0.25&11 (246)  & 2 (10)& 1.00 (0.56) & 0.18 (0.04)   \\
\hline
RT2.2\_D150z & 150 & $z$ & 1.68 &  89 (89)  & 2 (2)   & 0.12 (0.14) & 0.02 (0.02)   \\
RT2.2\_D300z & 300 & $z$ & 0.84 & 63 (110) & 11 (19) & 0.64 (0.49) & 0.17 (0.17)   \\
RT2.2\_D450z & 450 & $z$ & 0.56 & 34 (144) & 12 (33) & 0.85 (0.48) & 0.35 (0.23)   \\
RT2.2\_D650z & 650 & $z$ & 0.38 & 27 (211) & 11 (60) & 0.84 (0.38) & 0.40 (0.28) \\
RT2.2\_D860z & 860 & $z$ & 0.30 & 18 (258)  & 11 (90) & 0.69 (0.38) & 0.61 (0.35)   \\
RT2.2\_D1000z & 1000 & $z$ & 0.25&15 (280)  & 11 (114)& 0.68 (0.30) & 0.73 (0.41)   \\
\hline 
RT2.3\_D150z & 150 & $z$ & 1.68 &  135 (135) &  14 (14) & 0.50 (0.50) & 0.10 (0.10)  \\
RT2.3\_D300z & 300 & $z$ & 0.84 &  49 (105) &  16 (24) & 0.51 (0.27) & 0.32 (0.22)  \\
RT2.3\_D450z & 450 & $z$ & 0.56 &  44 (123) &  18 (37) & 0.58 (0.28) & 0.41 (0.30) \\
RT2.3\_D650z & 650 & $z$ & 0.38 &  37 (191) &  18 (61) & 0.58 (0.21) & 0.48 (0.32)  \\
RT2.3\_D860z & 860 & $z$ & 0.30 &  24 (178)  &  13 (98)& 0.42 (0.20) & 0.55 (0.54) \\
RT2.3\_D1000z & 1000 & $z$ & 0.25& 21 (287)  &  13 (147)& 0.41 (0.20) & 0.62 (0.51) 
\enddata
\tablecomments{For models RT2.1 - 2.3, ImSeg results outside parentheses are results for cores in the same physical region as the 150 pc distance model field, which covers a $\sim$2.3x2.3 pc region, while results in parentheses are cores in the whole field, including duplicates from tiling.}
\tablenotetext{a}{Resolution of 1 pixel before reprojecting}
\tablenotetext{b}{fraction of number of sink particles with cores to total number of sink particles in field}
\tablenotetext{c}{fraction of number of cores with sink particles to total number of cores}
\end{deluxetable*}

\begin{figure*}[tb]
\centering
\includegraphics[width=\linewidth]{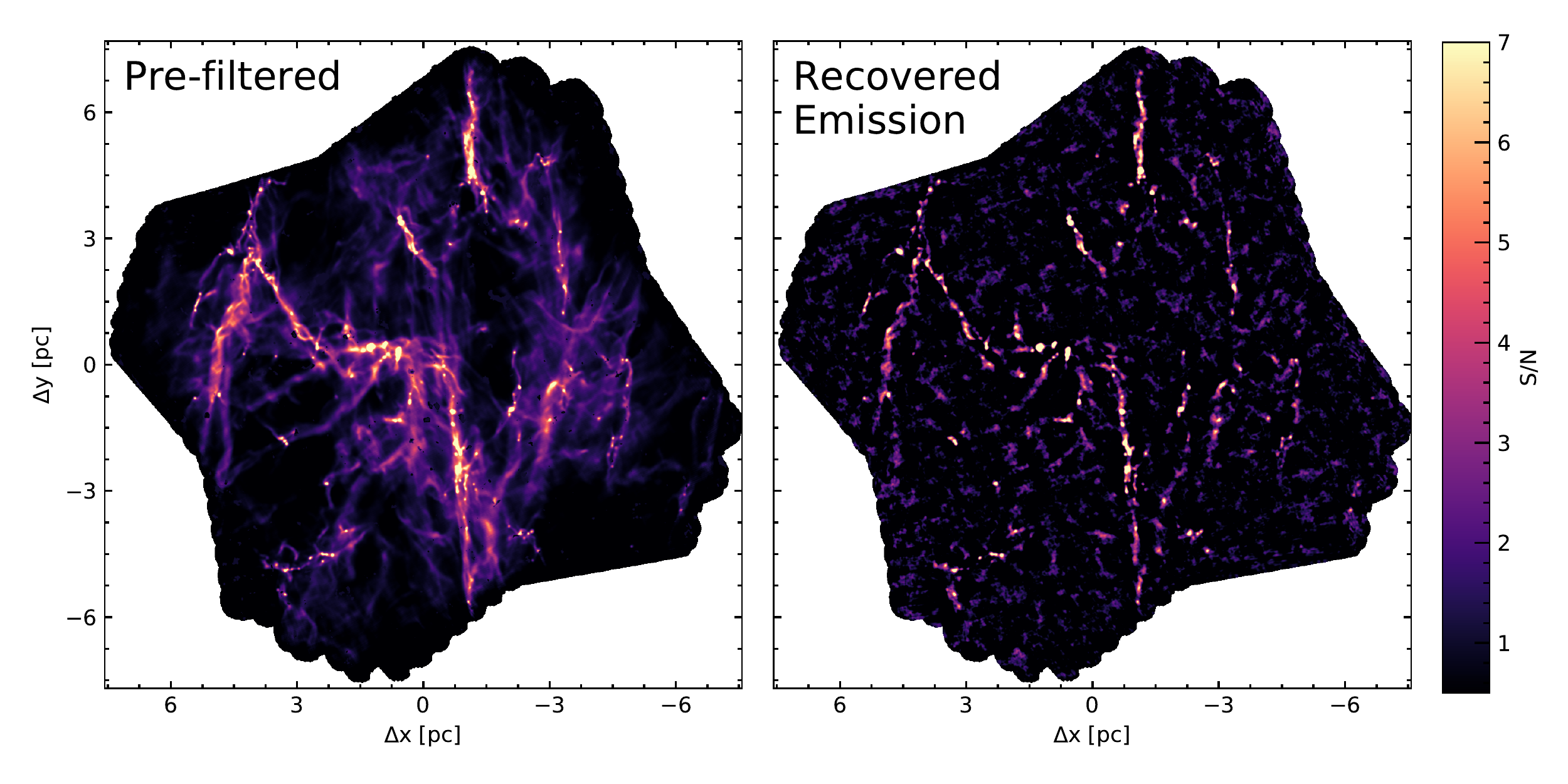}
 \caption{Pre-filtered synthetic observation S/N map (left) and post \texttt{macana} synthetic observation S/N map (right) for RT1$\_$D860z. }
 \label{simulation_pre_post_maps}
\end{figure*}

The scaled and beam-smoothed synthetic emission map is then added to the timestream data for Field~4 and reduced with \texttt{macana} to produce a synthetic emission map with the effects of atmospheric filtering.  However, in order to obtain a pure synthetic observation, the original AzTEC signal must be removed from the timestream.  This is achieved through the following steps:
\begin{enumerate}
    \item Reduce the original Field~4 timestream data to make a map of the Field~4 Mon~R2 cores.
    \item Negate the astronomical AzTEC emission (S/N $>$ 2.5) in the Field~4 map to create an inverted map.
    \item Project the negated astronomical signal Field~4 map along with the synthetic emission from the simulation back into the timestream to produce a synthetic observation of the simulation.
\end{enumerate}
During the reduction with \texttt{macana}, the negated AzTEC signal will be added to the original timestream, canceling it out and leaving only the synthetic signal.  However, removing the original emission in the timestream over-subtracts the signal, resulting in negative valleys at the locations of high S/N.  To get around this over-subtraction, we must only add a fraction of the negated signal back into the timestream, such that the final map consists of just the simulated signal. For Field~4, we find we must subtract off 55$\%$ of the original emission to produce a clean noise-only map.  See Appendix \ref{A1} for more details on how we determine this fractional amount.  Figure \ref{simulation_pre_post_maps} shows the pre-filtered synthetic observation map and the final synthetic flux map after running it through \texttt{macana} for RT2.3$\_$D860z.  Synthetic emission \texttt{macana} runs are listed in Table \ref{modelruns}.

%%% FIGURE 1

% -------  -------  --------------  -------  -------  -------  -------  -------  -------  -------  -------  -------  -------  -------  -------  -------  -------  -------  
\section{Core Property Extraction} \label{sec3}

\subsection{Core Identification} 
To identify the cores in the simulations, we create a python based core identification and selection process (ImSeg\footnote{\url{http://github.com/sbetti22/ImSeg}}) following the method described in \citet{Sokol2019}.  This method of core identification will be used by the C2C survey; therefore, we aim to test the robustness of this algorithm for detecting the same synthetic cores at various distances.

This method utilizes the python library \texttt{photutils.Segmentation} (p.Seg), a multi-thresholding algorithm for blob identification, which implements a watershed algorithm for separating blobs into individual components.  

Before identifying candidate cores, the signal map produced from \texttt{macana} is first masked on values with a S/N $<$ 2.5 and/or where the weight is less than 40 percent of the median non-zero weight values of the map.  This masking restricts our core search to those areas with relatively uniform coverage depth.  A noise map is also created from the signal and S/N map in order to give a threshold level for detection.  The masked signal map is then run through the p.Seg multi-thresholding algorithm.

Each segment found with the p.Seg multi-thresholding algorithm is run through a modified p.Seg deblending process.  This process utilizes the watershed algorithm to search for saddle points within each segment to separate an otherwise continuous emission region.  The resulting output is a footprint map of candidate dense gas cores. Properties, including total footprint area, center position, total and peak flux and S/N, and half-peak-power (HPP) information (position, peak and total flux, S/N, area) are calculated for each candidate core.  As in \citet{Sokol2019}, all cores that fall within 8 arcseconds of the coverage edge are subsequently rejected in order to eliminate the majority of false detections. 

In the \texttt{macana} reduction process, a jackknife technique is applied to the time stream data in order to characterize the noise.  For each reduction, 15 fully filtered and spatially mapped noise realizations are produced.  These noise realizations are passed through ImSeg and are used to estimate the false detection probability for each candidate core.  
To determine the final core candidates, we create 2D histograms of the total S/N and the ratio of the simulation derived column density to synthetic dust emission column density for both the noise realization false cores and core candidates.  The ratio of these two histograms is used to isolate regions of the parameter space where false detections dominate.  Confidence intervals are calculated for the ratio histogram.  If a core candidate lies within the 75$\%$ confidence interval, then it is considered a false detection.  Cores that lie outside this confidence interval have a low probability of being considered a false detection and are considered a core candidate.  

\subsection{Core Flux, Mass, and Size Measurements} 

\begin{figure}[tb]
\hspace{-0.5cm}
\includegraphics[width=1.1\linewidth]{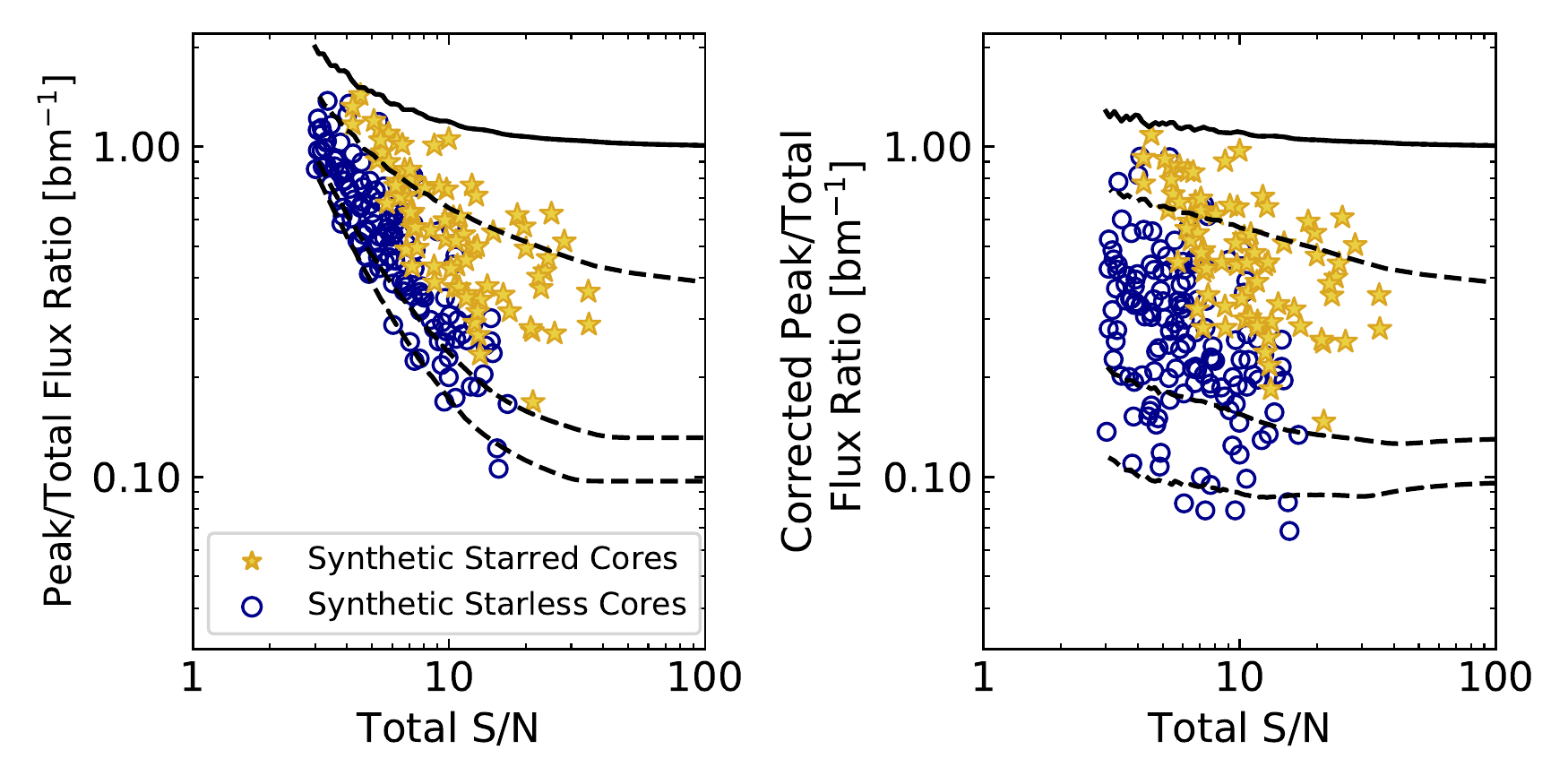}
 \caption{Peak to total flux ratio vs total S/N for all cores (blue-starless, gold-starred) in model run RT1\_D860z.  The left panel shows the uncorrected flux ratios with the ``iceberg effect" where lower S/N cores are not detected over their full extents.  The right panel shows the corrected flux ratio where the fluxes are not underestimated.  The final beam profile is shown as the solid line, while the Plummer models are shown as dashed lines.  The Plummer models have aspect ratios of 2, power law indices of 2, and scale lengths of 2, 12, and 18 arcsec (top to bottom).  }
 \label{flux_uncorr_corr}
\end{figure}

Once the synthetic cores are identified, we measure their sizes, masses, and temperatures.  As in \citet{Sokol2019}, we assume the synthetic cores are spatially resolved and the lower S/N cores are not detected across their full radial extents. \citet{Sokol2019} corrected the underestimation of total flux and core mass by modeling and characterizing the peak to total flux ratio relation and then correcting for high ratios at low S/N.  They tested their correction by constructing several Plummer-like models that span the expected ranges of the peak-to-total flux and total S/N of pre-stellar cores.  To confirm our simulated cores have these same features and we can use the same correction, we model the peak to total flux ratio relation and find the same ``iceberg" effect seen by \citet[][see their Figure 7]{Sokol2019} in all model runs.  We then apply the same correction 
\begin{equation}
F_\mathrm{corr} = F_\mathrm{peak} \left(\frac{F_\mathrm{peak}}{F_\mathrm{tot}} - \delta R \right)^{-1}
\end{equation}
where $F_\mathrm{peak}$ is the observed peak flux, $F_\mathrm{tot}$ is the observed total flux, $F_\mathrm{corr}$ is the corrected total flux, and $\delta R$ is the S/N dependent flux correction term found to be
\begin{equation}
\delta R = 5.25 \times (S/N)^{-1.8}.
\end{equation}
The simulated core flux ratios are corrected for the noise bias.   
We show fluxes before and after correction for RT1\_D860z in Figure \ref{flux_uncorr_corr}.  We use this process to correct all total observed fluxes for all model runs.

%%% FIGURE 2

We then calculate the mass of the cores from the corrected fluxes.  At 1.1~mm, the mass is found to be
\begin{equation} \label{Mass}
\begin{split}
M = 1.3\times 10^{-5}\left(\frac{S_{1.1\ \mathrm{mm}}}{1 \ \mathrm{Jy}}\right) \left(\mathrm{exp}\left(\frac{13 \ \rm K}{T_d}\right)-1 \right) \\
\times \left(\frac{D}{1\ \rm pc} \right)^2  \left(\frac{\kappa_{1.1\ \rm mm}}{0.0121 \mathrm{\ cm^2\ g^{-1}}}\right)^{-1} \rm M_\odot
\end{split}
\end{equation}
where $S_{1.1\ \rm mm}$ is the total flux density at 1.1 mm, $T_d$ is the dust temperature, D is the distance, and $\kappa_{1.1\ \rm mm}$ is the dust opacity at 1.1 mm.  Following \citet{Sokol2019}, we take the dust opacity to be 0.0121 cm$^2\ \rm g^{-1}$ at 1.1 mm.  The dust opacity is found from the \citet{Ossenkopf1994} model 4 opacities for icy dust grains at 1.1 mm, and assumes a gas-to-dust ratio of 100.  

Core sizes are calculated following \citet{Sokol2019} and \citet{Konyves2015}.  The widths of the cores are found from the deconvolved FWHM size given by 
\begin{equation}
\rm size_{\rm deconv} = \sqrt{FWHM^2 - HPBW^2}
\end{equation}
where the FWHM is the half peak power diameter and the HPBW is the final AzTEC beam width (12$\arcsec$).  Like in \citet{Sokol2019}, there is a bias toward smaller areas for cores with a S/N $<$ 5.  We therefore use the correction found by \citet{Sokol2019} to calculate the unbiased half peak power areas of the cores in order to determine the FWHM.

The temperature of each core is found from temperature maps derived from the simulated radiation energy density smoothed to the \textit{Herschel} beam at 500 $\mu$m (36$\arcsec$) at each model distance.  We take the average temperature within the footprint of each core as the temperature.  As the \textit{Herschel} beam and pixel scale is larger than the AzTEC maps, the synthetic temperature maps may be biased due to blending and the large beam causing temperatures to be overestimated.  Since this same effect will occur for the C2C survey as \textit{Herschel} temperature maps will be used for analysis, this provides a good test for temperature robustness.  

%----------------------------------------------------
\section{Core Property Systematic Effects Assessment} \label{sec4}

With synthetic observations, we can probe how measurements of identical molecular gas structures are impacted by observational and analytical algorithm biases. Here we describe two systematic biases identified by placing the simulations at a range of heliocentric distances.

\begin{figure}[tb]
\centering
\includegraphics[width=\linewidth]{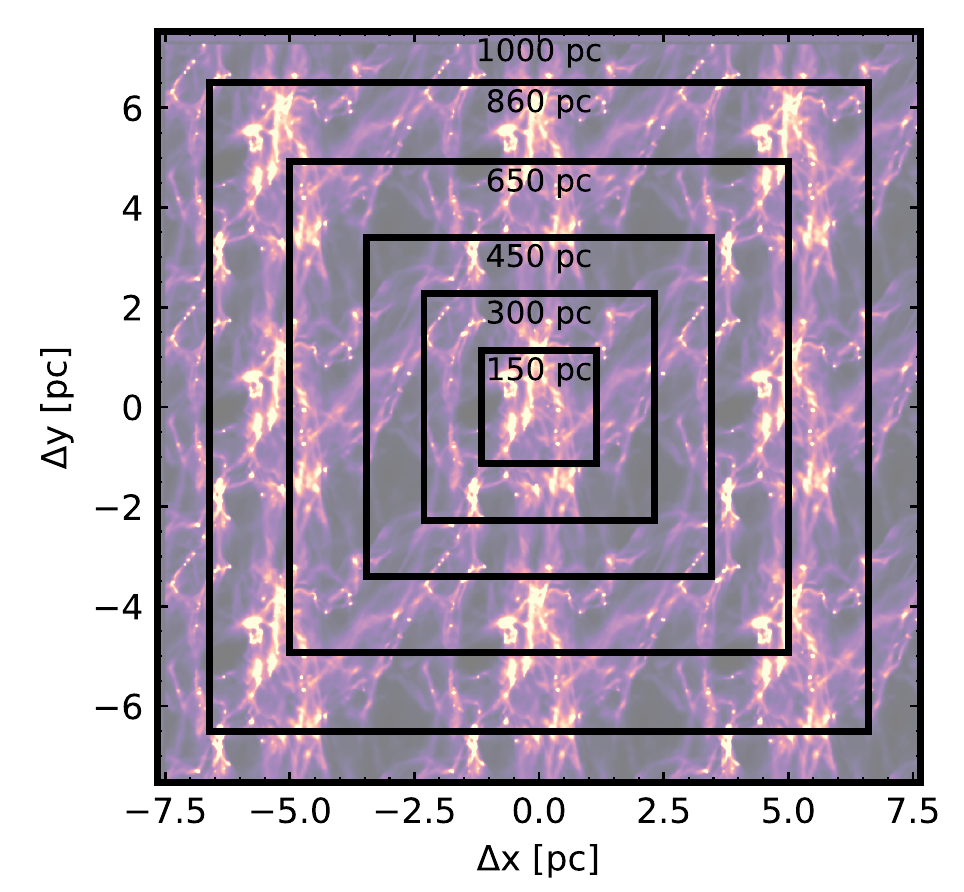}
 \caption{RT2.3 overlaid with the sizes of the simulation at different distances (black boxes).  At different distances, different amounts of the simulation fit into the Field~4 area.}
 \label{distance_coverage}
\end{figure}

Using the three RT2 snapshots, we look at how varying the distance and age of the cloud affects core properties.  This will help place constraints on the CMFs and core sizes from clouds at various distances and ages that are planned to be observed in C2C.  The three RT2 outputs are placed at 150, 300, 450, 650, 860, and 1000~pc in order to cover the full range of distances in the C2C survey.  860~pc is specifically chosen to directly compare with the cores from Mon~R2 from \citet{Sokol2019}.  In Figure \ref{distance_coverage}, the black boxes indicate the extent of the tiled pre-filtered synthetic observations seen at each distance.  RT2.3 is tiled and shown underneath for visualization purposes.

%% FIGURE 3

At 150~pc, only the center 2.3~pc of the synthetic emission map fits within the Field~4 coverage; therefore, when comparing cores at all distances, we only look at cores within the center 2.3~pc of each model to make a fair comparison.  This region will be called the ``center", while the whole Field~4 coverage will be called ``total".  The center core candidates found for each model run are summarized in Table \ref{modelruns} along with total core candidates shown in parentheses.  On average, for all synthetic emission model runs, we recover cores subtending 65$\%$ of sink particles within the center and $\sim 60\%$ of all sink particles.  

%% FIGURE 4
\begin{figure}[tb]
\centering 
 \includegraphics[width=\linewidth]{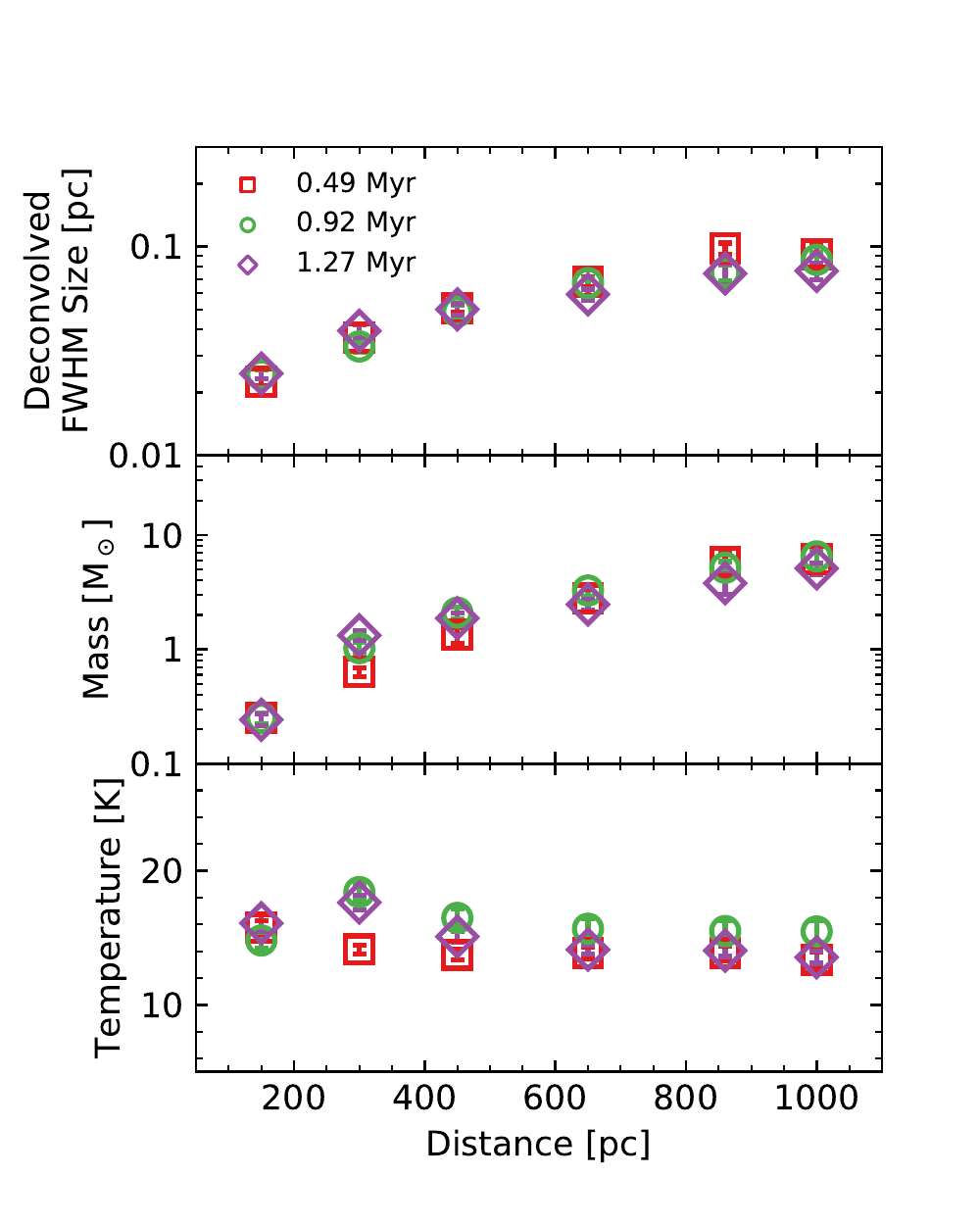}
  \caption{Median recovered sizes, masses, and temperatures as a function of distance for starless and starred cores. For all panels, red squares are RT2.1, green circles are RT2.2, and purple diamonds are RT2.3.  Cores show a size and mass increase at large distances, indicative of over-segmentation and over-filtering at the closer distances.}
  \label{sim_medians}
\end{figure}

We find that the median recovered masses, sizes, and temperatures for the synthetic cores vary with distance for the three RT2 synthetic observations, as shown in Figure \ref{sim_medians} (red is RT2.1, green is RT2.2, and purple is RT2.3).  From 150 to 1000~pc, the mass increases by 1.4~dex, the size increases by 0.64~dex, and the temperature decreases by $37\pm6\%$. At D~$>$~450~pc, these physical properties do level off, varying only by $\sim$0.2~dex.  Though there is slight temperature dependency with distance, the order of magnitude discrepancy in mass cannot be fully explained by this.  If the temperature was the only cause of the mass discrepancy, the temperature would have to vary by $\sim$1~dex in the opposite sign to our result, with core temperatures around 6~K at 150~pc and 60~K at 1000~pc.  
As this is not the case, the discrepancy in mass and size is not due to the variation in temperature.  

This significant increase in core mass and size with distance is problematic as these properties should be intrinsic quantities and not dependent on distance. Though we could be probing substructure within the cores themselves, we clearly need to identify and extract similar gas structures across all distances to ensure their fair comparison. 
We consider two likely causes of the observed discrepancies here: atmospheric filtering in our data reduction process and over-segmentation during the core identification and extraction analysis. The former is likely to shrink the extent of diffuse emission for clouds at closer distances. The latter occurs when resolved substructures within cores are considered distinct objects at close distances and yet are blended together at further distances. 

\begin{figure*}[tb]
\centering
\includegraphics[width=.8\linewidth]{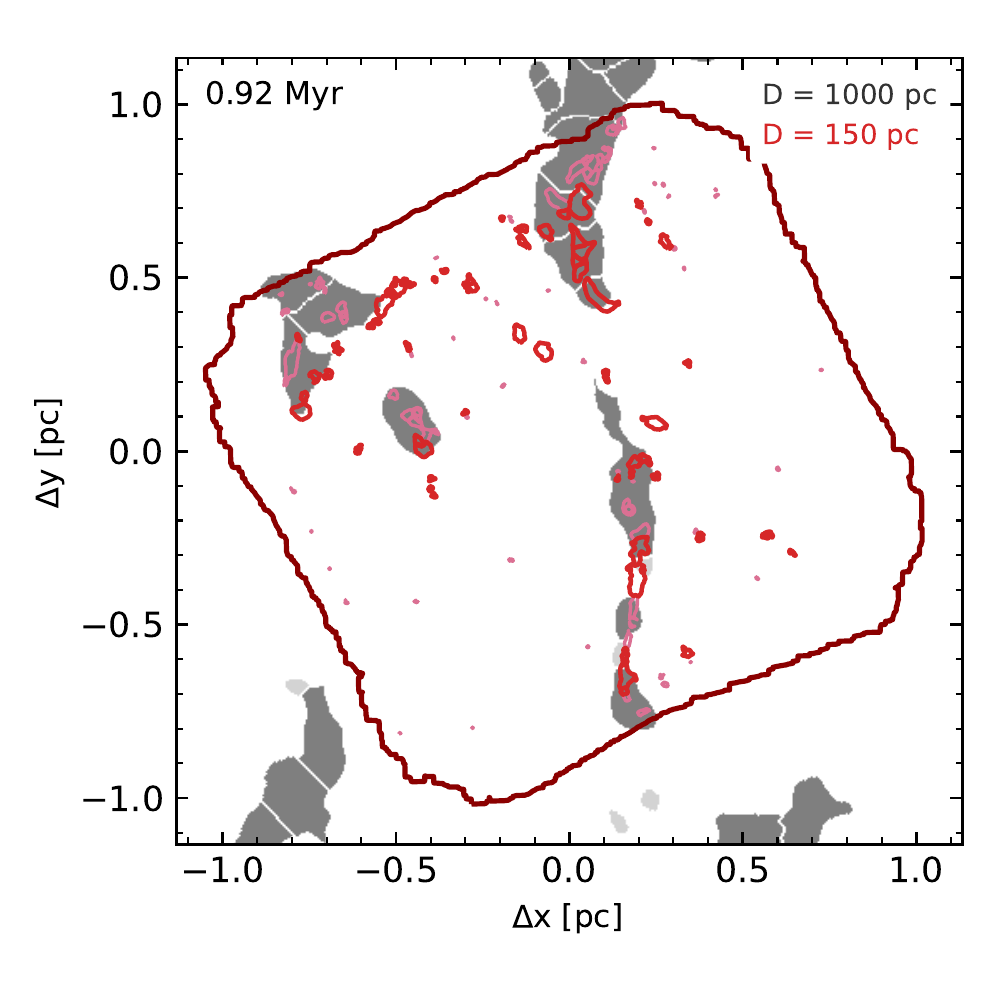}
  \caption{Footprint map at 1000~pc (grays) overlaid with the 150 pc contours (reds) for RT2.2.  False core detections are indicated by light gray/pink, while core candidates are shown as dark gray/red. The dark red outline is the coverage edge for the 150~pc map.  The over-segmentation and over-filtering at 150~pc is shown by the increased number and narrower area of the segments compared to the cores at 1000~pc.}
  \label{over-filtering}
\end{figure*}

We present an example comparison in Figure~\ref{over-filtering}, where the red/pink contours show the footprint map for segments at 150~pc overlaid on gray footprint maps of segments at 1000~pc.  The pink and light gray contours are false core detections, while the red and dark gray are cores detections.  The most obvious difference is that there are several smaller segments at 150~pc for every segment found at 1000~pc. Using the ratio of core counts in Table~\ref{modelruns}, there are $\sim$6 times as many cores at 150~pc compared to 1000~pc, though of course there are other minor differences in the cores found at each distance. This segmentation discrepancy results in the closer distance extraction providing many more segments overall, and with commensurately smaller typical masses and sizes. The apparent ``cores'' in the 150~pc synthetic observation have an average size of 0.025~pc (5100~AU) and mass of 0.25~M$_\odot$, similar to the protostellar ``envelopes'' found {\it within} cores by mm-wave interferometers \citep{Pokhrel2018}. Regardless of their origin, extracting only the smallest resolvable structures will not yield a consistent set of structure properties across our distance range.  

%% FIGURE 5

\begin{figure}[tb]
\centering
\includegraphics[width=.8\linewidth]{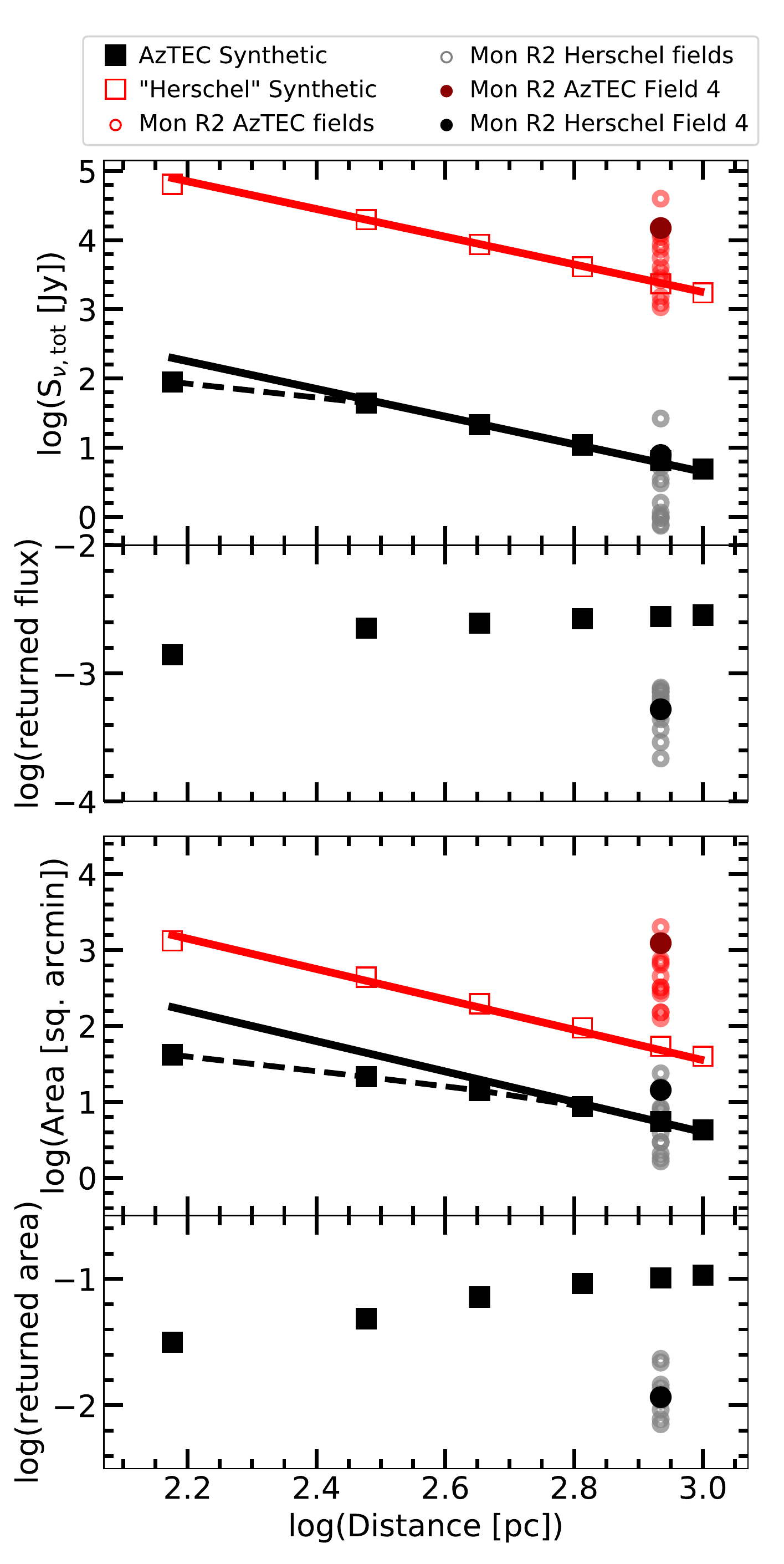}
  \caption{\textit{top} - Total flux and total area of the synthetic observations and Mon~R2 AzTEC cores. \textit{bottom} - Percent of flux and area recovered as a function of distance.  The synthetic observations are shown as black squares and the ``\textit{Herschel}-like'' 36\arcsec 1.1~mm pre-filtered map are red squares within the center 2.3~$\times$~2.3~pc region.  The gray circles are the AzTEC Mon~R2 total flux and area within each field mapped by AzTEC while the black circle is Field~4 \citep{Sokol2019}.  The red circles are the \textit{Herschel} total flux and area within each AzTEC field and the dark red circle is Field~4.  The solid black and red lines follow the $D^{-2}$ dependence for flux and area, while the dashed black lines show the deviation from this dependence. The uncertainties in flux and area are insignificant for all maps.}
  \label{flux_area_residuals}
\end{figure}

The combined red 150~pc footprint is also distinctly narrower than the footprint at 1000~pc due to strong negative halos surrounding the positive flux, a classic sign of atmospheric filtering affecting the detectable astrophysical signal. If over-filtering was not an issue, the total amount of flux and the emission area within the same physical area should scale inversely with distance squared.  However, as shown in Figure \ref{flux_area_residuals}, there is a substantial loss of flux and area at relatively close distances. At 150~pc, only 1.3$\times10^{-3}$ of the total original flux is recovered compared to $\sim$3$\times10^{-3}$ that is recovered at larger distances, a factor of 2.3 greater. The emission footprint total area (pixels greater than the median noise, $\sim$7~mJy) is similarly discrepant, with only 0.03 of the total original area recovered at 150~pc compared to 0.1 of the total area recovered at 650$-$1000~pc (uncertainties for both flux and area are insignificant here), a factor of 3.3 greater.  This small reduction of region-wide mean flux density within the identified emission footprint, by a factor of 0.7, despite a net increase in total flux detected masks complex flux-filtering behavior. Ultimately, the median mass surface densities of cores increased by a factor of 2.3 over our distance range. This is opposite the sign of the change in mean flux density, but consistent with the increase in total flux within the fixed field of view of the center region (the median core temperature difference also impacts the mass to flux conversion).  

In summary, the mass and size discrepancies of cores at close distances are a direct result of both over-filtering and over-segmentation.  
Care should be taken in the C2C survey when comparing cores at various distances, and future work will need to address these issues, as the smaller TolTEC beam may exacerbate the issues at the closest distances. With that stated, it is expected that TolTEC data will not suffer as much filtering damage because of its larger array footprint and massive increase in sampling density in space (35 times more detectors at 1.1~mm) and time (higher sampling frequency) over AzTEC. Furthermore, TolTEC's simultaneous observations at three mm wavelengths may enable additional advances in atmospheric filtering that are less destructive to astronomical signal.  However, for this work, we will only use synthetic observations at D $>$500~pc for the remainder of the analysis so that the effects of these biases are negligible.

%%% FIGURE 6

\section{Comparison to AzTEC Mon~R2 Observations} \label{sec5}

To analyze how well the simulations can reproduce observed core properties within molecular clouds, we compare our synthetic cores to observed cores in Mon~R2 \citep{Sokol2019}.  We use simulation RT1 as it is approximately the same size as Field~4 (10~pc simulation vs $\sim$12~pc AzTEC Mon~R2 field) and therefore the simulation fills the whole field with little duplication from tiling.  At this distance and scale, sink particles are proxies for YSOs.  The core candidates found for model run RT1$\_$D860z are summarized in the first row of Table \ref{modelruns}.  All cores where a sink particle falls within the core footprint are considered starred, while all other cores are considered starless.  We find 222 core candidates with 35$\%$ of them considered starred.  However, only $\sim 53\%$ of all sink particles are located within the footprint of a core.  The sink particles not within core footprints are in areas of low S/N emission and are thus not detected in the synthetic observation.   

\begin{figure}[tb]
\centering
\includegraphics[width=\linewidth]{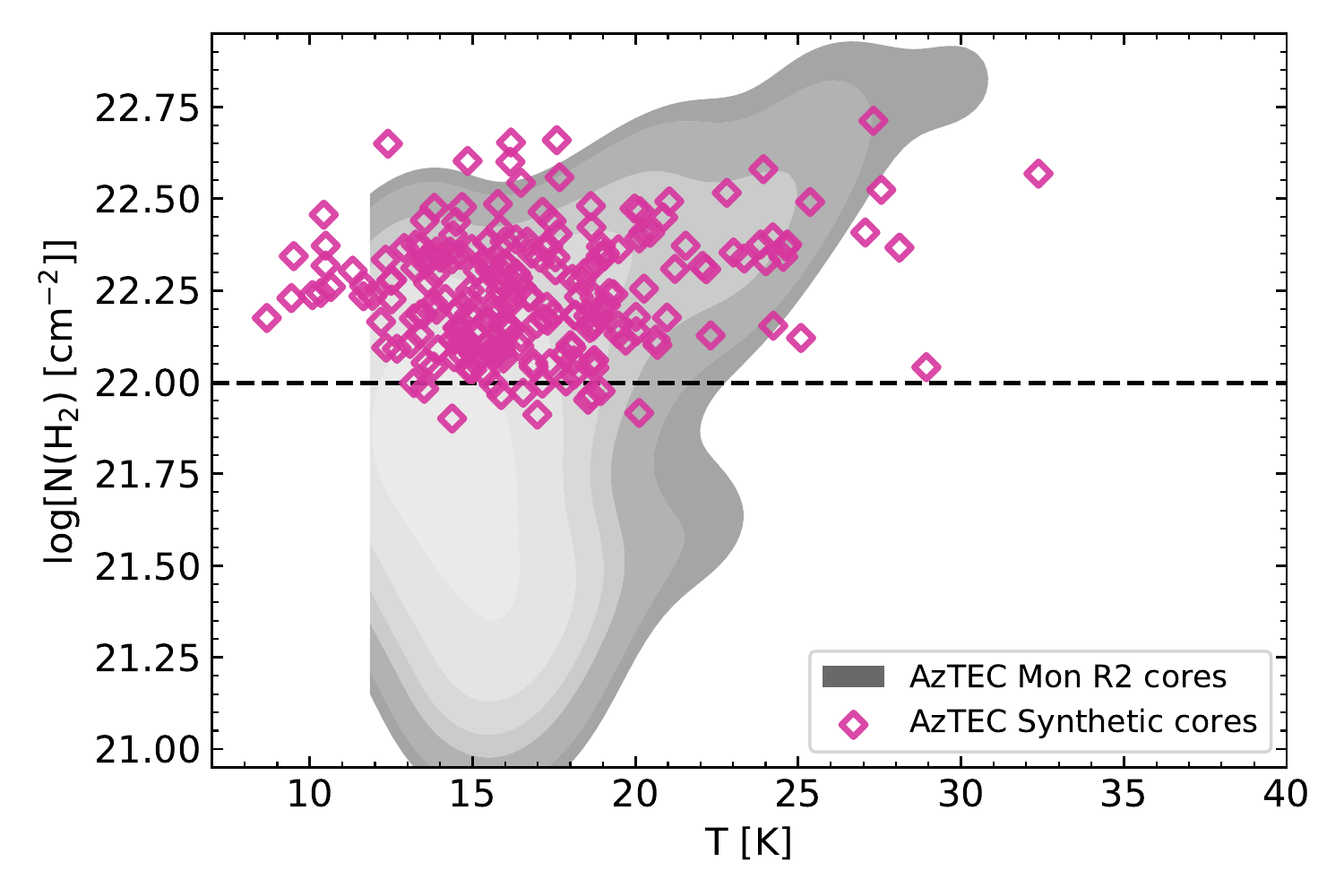}
 \caption{Column density vs temperature for cores from the RT1\_D860z synthetic observation (magenta diamonds) and AzTEC Mon~R2 observations (gray contour).  In order to make a fair comparison between synthetic and real observations, we want a sample of cores that lie in the same type of environments.  Therefore, we only select observed cores with column densities N(H$_2) > 10^{22}$ cm$^{-2}$ (which is the 5$^{\rm th}$ percentile synthetic column density value; black dashed line.)   }
 \label{pltsn08550_NH2}
\end{figure}

The simulations generally best represent conditions found in denser ``clump'' regions within nearby molecular clouds (e.g. the center of Mon~R2). Thus we must attempt to prune the observed cores to those found in similar environments to the simulation.  We select observed cores with local diffuse gas column densities N(H$_2) > 10^{22}$ cm$^{-2}$ (the 5$^{\rm th}$ percentile simulated column density value) as shown in Figure \ref{pltsn08550_NH2}.  All cores with N(H$_2) > 10^{22}$ cm$^{-2}$ are relatively dense but span a wide range of temperatures and fill the same parameter space as the synthetic cores.  Only observed cores above this column density will be used in order to make a fair comparison of the properties of the two sets. 

%% FIGURE 7

%% FIGURE 8
\begin{figure*}[tb]
\centering
\includegraphics[width=\linewidth]{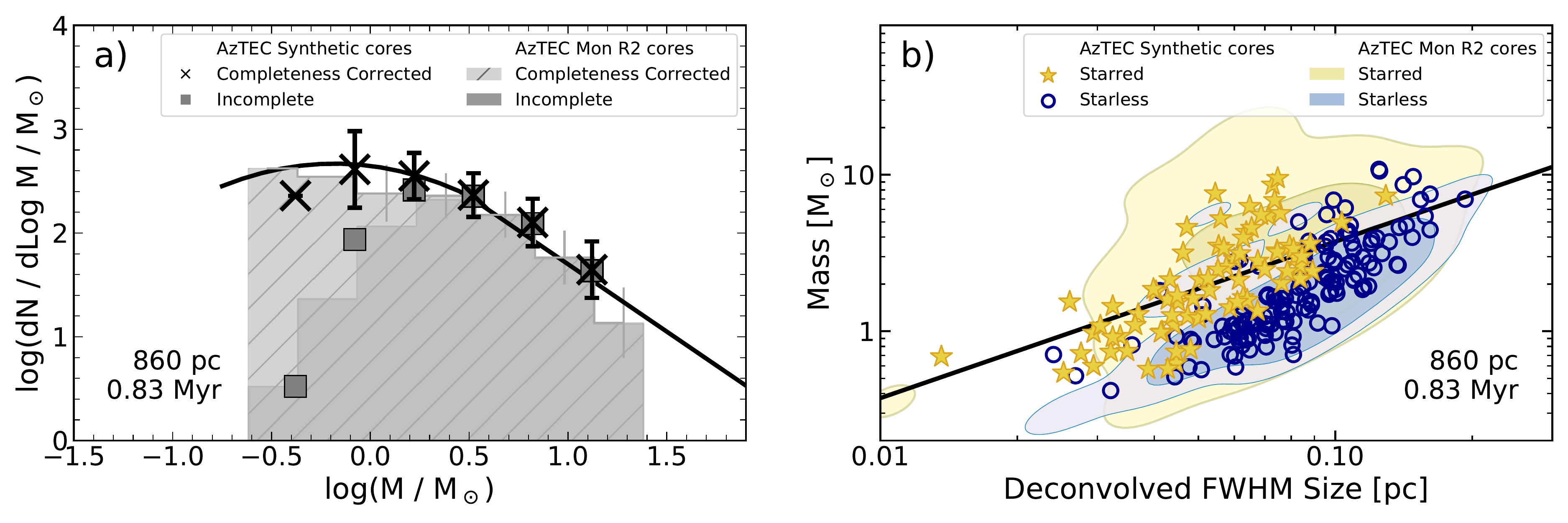}
 \caption{RT1\_D860z synthetic core observations compared to Mon~R2 AzTEC Field~4 cores with similar column densities.  \textit{a)} RT1\_D860z synthetic core mass function (x/square) and AzTEC Mon~R2 core mass function (histograms).  The incomplete CMF is shown as squares/dark histogram, while after correcting for completeness is shown in x/hatched histogram.  Error bars are shown in the corrected CMFs if they have a $>$~20$\%$ completeness.  A \citet{Chabrier2003} IMF with a mass scale factor of 3 is shown as the black line. \textit{b)}
 RT1\_D860z synthetic (blue circles/gold stars) and AzTEC Mon~R2 Field~4 (blue/gold contours) corrected mass vs corrected deconvolved FWHM size.  Cores with YSOs for the synthetic observation and the AzTEC data are shown as gold stars and contours, while starless cores are blue circles and contours.  The black line is the Bonnor-Ebert stability line for the median core temperature \citep[T=16~K;][]{Konyves2015}.  Cores above this line are gravitationally bound, while cores below this line are gravitationally unbound.
 }
 \label{pltsn08550_cmf_MS}
\end{figure*}

In order to determine the extent we can accurately compare synthetic cores, and therefore the robustness of the simulations in predicting core properties and environmental effects, we first look at the core mass functions (CMF) for the synthetic cores ($\times$/square) and AzTEC Mon~R2 observations (histograms) in Figure \ref{pltsn08550_cmf_MS}a.  As the synthetic observations are made using the same spatially filtered maps and reduction process, we use the same differential core detection completeness characterization and corrections as in \citet{Sokol2019}.  In that work, the authors inserted nine false cores per mass bin into the timestream data, reduced the modified data with \texttt{macana}, and ran a full core extraction analysis to determine the differential completeness as a function of corrected core flux (see their Figure~6).  Since we use their data for our synthetic observations, we also adopt their completeness trend to correct for the incomplete core number counts at the low mass end of our synthetic core mass function. % by dividing it into the CMF.    

We find that the CMF shape derived from the synthetic emission cores is the same as that of \citet{Sokol2019} with and without completeness corrections.  The mass complete values in Figure \ref{pltsn08550_cmf_MS}a are shown as $\times$'s/light gray hatched histogram, while the mass incomplete observed values are gray squares/gray hatched histogram.  The number counts between the starless and starred synthetic cores vary, with 175 starless synthetic cores and 83 starred synthetic cores.  
We cannot rule out a turnover at the low mass end near 3~M$_\odot$, similar to a Chabrier IMF with a mass shift of 3$\times$ (black line).

Figure \ref{pltsn08550_cmf_MS}b shows the mass and sizes of the synthetic cores separated as starless (blue unfilled circles) and starred (yellow filled stars) overlaid on the parameter space occupied by Mon~R2 cores (pre-stellar cores as yellow contours, and starless as blue contours).  The median synthetic core size is 0.073~pc while the median Mon~R2 core size is 0.083~pc.  We see a clear trend of small cores (cores with sizes $< 0.05$~pc) containing primarily sink particles not seen in observations.  In RT1$\_$D860z, separated at the median size (0.073~pc), 51 $\pm \ 7\%$ of cores $<$~0.073~pc contain sink particles compared to 23~$\pm \ 4\%$ of cores $>$ 0.073~pc. This substantial difference among core sizes is not observed in the Mon~R2 cores, where 33~$\pm \ 5\%$ of cores with sizes $<$~0.083~pc contain YSOs compared to 23~$\pm \ 7\%$ for larger sizes. This discrepancy appears for two reasons. First, inspection of the raw simulation data (see Appendix~\ref{App3}) shows that these sources are mainly older objects, which have a bright, massive disk and very little surrounding envelope. Hydrodynamic simulations, which neglect magnetic fields, commonly produce large disks \citep{Zhao2020}, which are much more massive than expected compared to observations \citep[e.g.,][]{Williams2019,Tobin2020}. On the observational side, the region of the parameter space for smaller low-mass cores, $M \lesssim 0.5$~M$_{\odot}$, has a completeness of $< 5\%$, so we expect more starless and protostellar objects to reside in this region than are actually detected.

In order to gauge how environment may affect the cores found in our synthetic observations, we examine how the cores are clustered relative to the column density of their surrounding diffuse gas.  Following \citet{Sokol2019}, surface density is calculated by finding the nearest neighbor distances (d$_n$) from each core given as ($n$-1)/($\pi$(d$_n$)$^2$).  This is multiplied by the mean mass of the $n$ cores selected to get the core mass density.  Using our $Herschel$-like synthetic column density map, we find the average column density over the same area used to calculate the core mass density.  We measure the gas surface and core surface densities for $n =$~4, 6, 11, and 18 nearest neighbors to look at clustering at different size scales, similar to \citet{Gutermuth2008}, \citet{Sokol2019}, and \citet{Pokhrel2020}.  
 
\begin{figure}[tb]
\centering
\includegraphics[width=\linewidth]{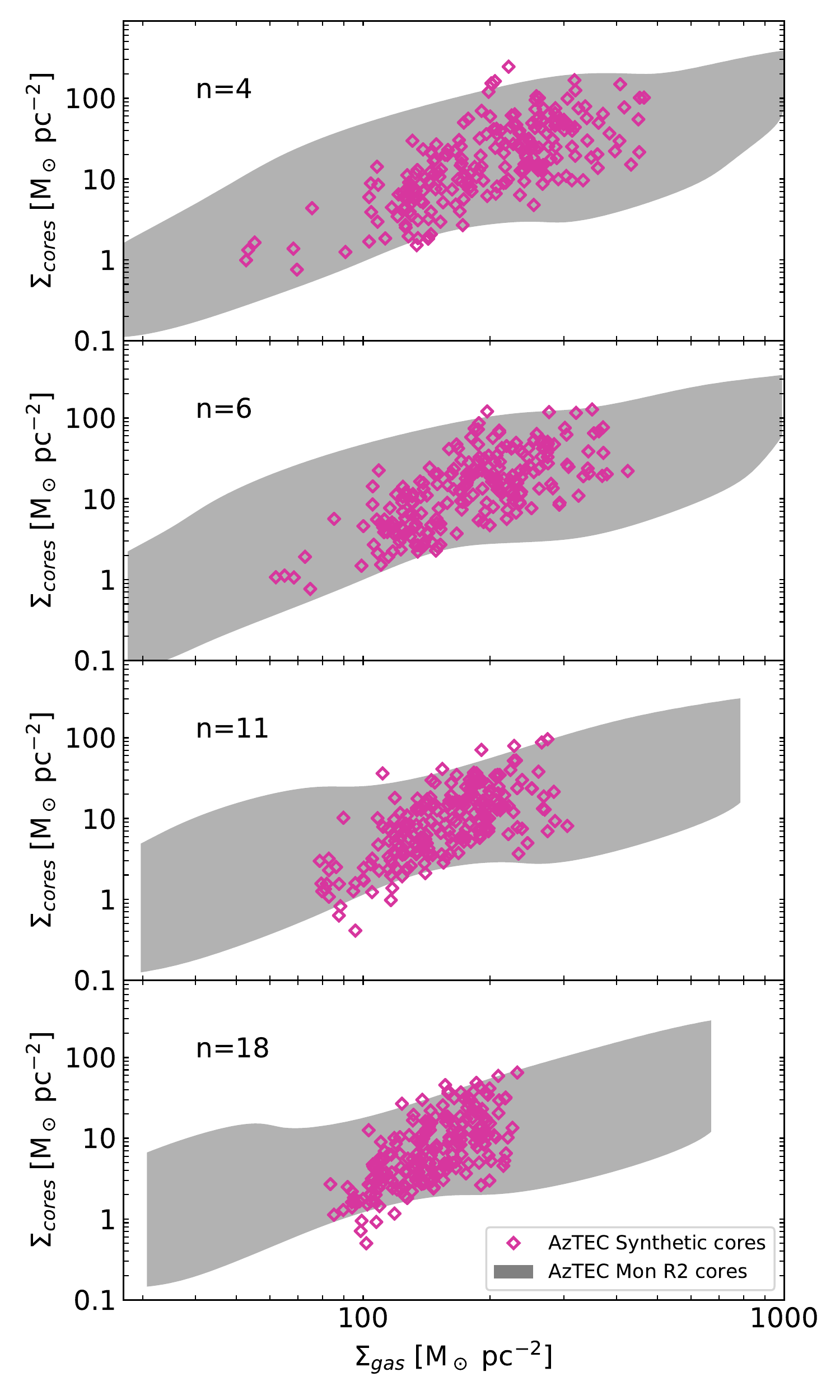}
 \caption{RT1\_D860z synthetic observation (magenta) and  AzTEC/\textit{Herschel} Mon~R2 (gray contour) core-gas correlation for $n = $4, 6, 11, and 18 nearest neighbors.  At high gas and core surface densities, cores are clustered within small dense regions, while at lower surface densities, the cores are in a more diffuse medium.  As $n$ increases, the amount of synthetic clustering in dense regions decreases.}
 \label{pltsn08550_CGC}
\end{figure}

\begin{figure}[tb]
\centering
\includegraphics[width=\linewidth]{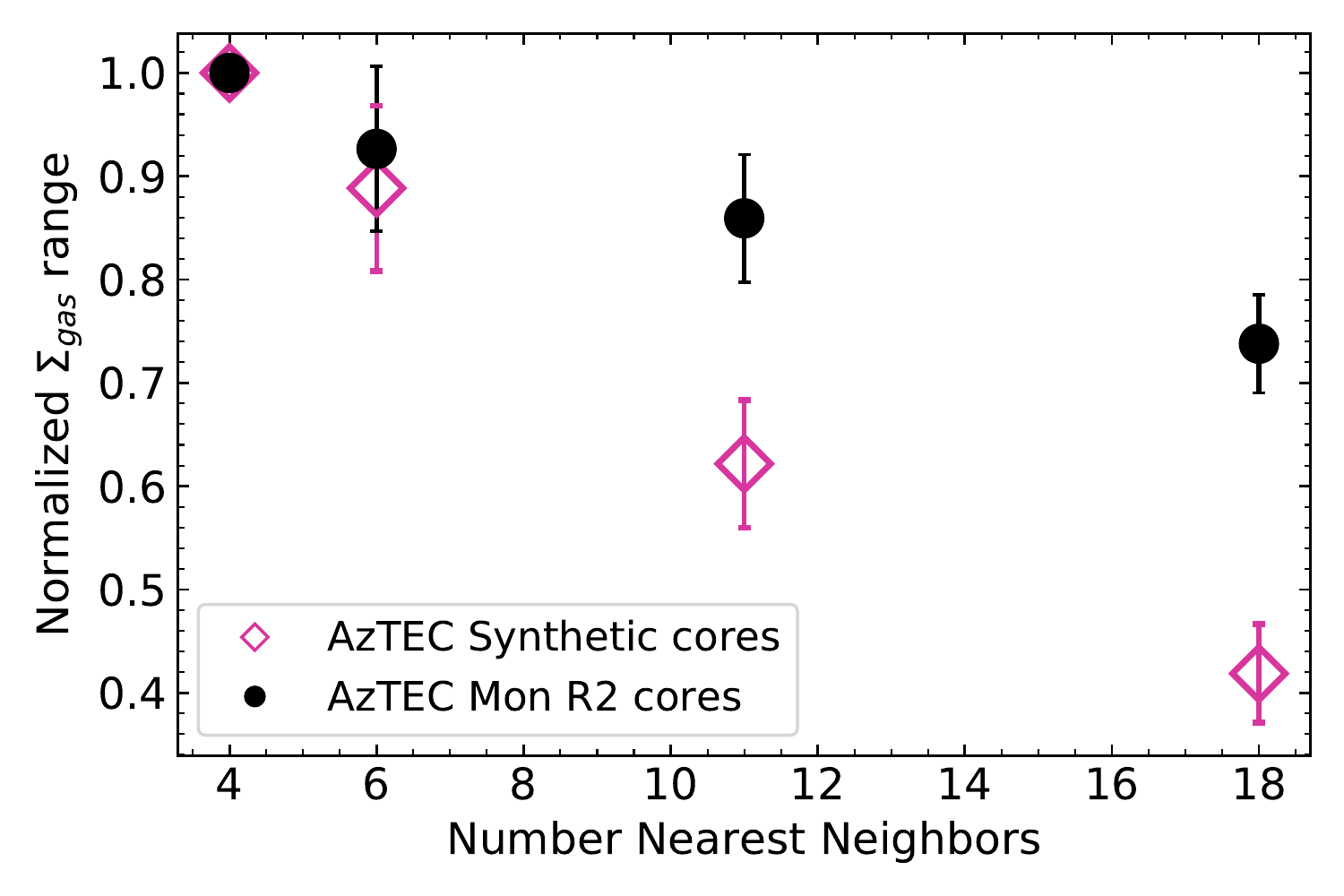}
 \caption{RT1\_D860z synthetic observation (magenta) and AzTEC/\textit{Herschel} Mon~R2 (black) gas surface density ranges from Figure \ref{pltsn08550_CGC} normalized to the $n=4$ range.  As $n$ increases, the range in synthetic gas surface densities decreases at a faster rate than observations.}
 \label{pltsn08550_CGC_avg}
\end{figure}

As shown in Figure~\ref{pltsn08550_CGC}, RT1 (pink diamonds) tends to span a smaller range of gas densities than the Mon~R2 observations, and that range decreases as more neighbors ($n$) are included.  This effect is largely confined to the high column density end, while the the minimum gas densities remain relatively constant.  This indicates more diffuse gas (lower column density) is enclosed as the smoothing size scale increases and the enclosed area also increases in order to contain more nearest neighbors.  Figure \ref{pltsn08550_CGC_avg} shows that when we compare the range of gas densities subtended, we see much stronger dilution for a given $n$ compared to the Mon R2 observations.  The synthetic gas densities shrink by $59\%$ by $n=18$, while the Mon~R2 cores only shrink by $27\%$ over the same range in $n$ value selection. We also find the average gas surface density in the synthetic observations shifts by a greater extent as more neighbors are included, from 212~M$_\odot$ pc$^{-2}$ to 148~M$_\odot$ pc$^{-2}$, while the Mon~R2 observations shift from 210~M$_\odot$ pc$^{-2}$ to 171~M$_\odot$ pc$^{-2}$.   This shrinking and shifting is predominantly a signature of smaller-$N$ core groupings in the simulation than in Mon~R2, a systematic difference between the two data sets.

% FIGURE 9

% FIGURE 10

Overall, while the simulation can accurately reproduce the observed Mon~R2 CMF for cores in a similar range of column densities, the environments that form these cores differ from observations.  While some cores form in areas of low column densities, the majority of low mass cores preferentially form in small but overly dense regions.  The discrepancies in clustering characteristics between the observations and synthetic results demonstrates the limiting nature of the simulations used here for disentangling environmental effects on core properties. Future work will employ next generation simulations, such as those from the STARFORGE project \citep{grudic2020}, that simulate entire molecular clouds over a much wider dynamic range of spatial scales and gas column densities and incorporate more stellar feedback effects.  Parallel improvements in the simulation suite used and the observations from the TolTEC Clouds to Cores survey should facilitate much more comprehensive analysis of environmental impacts on core properties.             

\section{Core Property Evolution from Multiple Simulation Snapshots in Time} \label{sec6}

In order to analyze cores across various environments and ages from the C2C survey, we first need to characterize systematic effects in the data reduction and analysis process, determine the extent simulations can reproduce observations, and if the simulations are robust, start to probe how age will affect observed core properties.

Understanding both the systematics in the data reduction techniques and the reliability of simulations is essential for explaining and interpreting observations.  Therefore, we focus our analysis on both these angles with the goal of characterizing potential issues that will arise in the C2C survey results and interpretation.  We will show the differences between synthetic cores compared to observed core properties highlighting potential issues with our understanding of star formation within simulations.

\subsection{Mass and Size}
%% FIGURE 11
\begin{figure*}[tb]
\centering
\includegraphics[width=\linewidth]{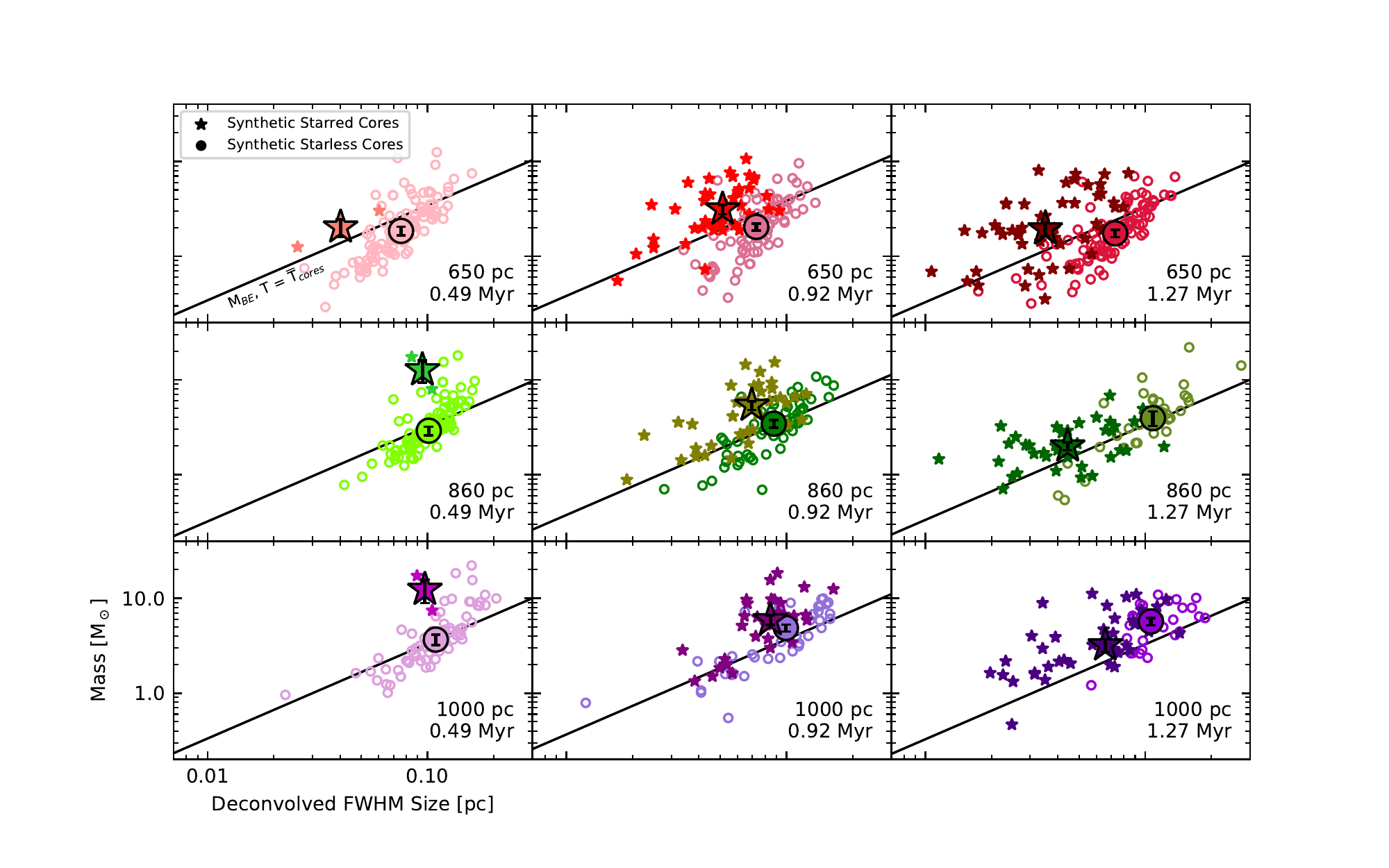}
 \caption{Corrected mass vs corrected deconvolved FWHM size for RT2.1 (left), RT2.2 (center), and RT2.3 (right) at 650 (top), 860 (middle), and 1000~pc (bottom) within the center 5 pc of each model.  The darker filled stars correspond to the starred cores, while the lighter unfilled circles are starless cores.  The large star or circle is the median value for the starred or starless cores, respectively.  The black line is the Bonner-Ebert stability line for the median core temperature \citep[$\approx$15~K for all panels;][]{Konyves2015}.   By 1.27~Myr, there is a split in size for starred and starless cores.}
 \label{NEW_mass_size}
\end{figure*}

We first characterize how the inferred masses and sizes (and therefore the CMF) will vary with both distance and age, two of the most prominent dependent variables for the C2C survey sample. 
Overall, the starless core properties are independent of age, with little variation in average mass, size, and temperature with time (Figure \ref{NEW_mass_size}).  However, as the cloud ages, starred cores appear above the distribution of starless cores, suggesting that cores grow in mass before undergoing collapse. Over time the starred cores, on average, collapse and shrink, from a median mass of 12~M$_\odot$ to 2~M$_\odot$, and size of 0.98~pc to 0.04~pc. This represents the depletion of the envelope as the protostar accretes.

In Figure \ref{NEW_mass_size}, we show the mass-size relation for 650, 860, and 1000 pc at all ages with starred cores shown as stars, and starless shown as circles.  The black line is the approximate numerically modelled Bonnor–Ebert (BE) stability criterion line \citep{Konyves2015}.  The critical BE mass is the largest mass an isothermal sphere in a pressurized medium can have while keeping hydrostatic equilibrium \citep{Bonnor1956}.  A core is considered self-gravitating and bound if its mass is above the approximate numerically modeled critical thermal BE mass, $M_{\rm BE, crit} = 2.4 R_{\rm BE} c^2_s / G$.  $R_{\rm BE}$ is the BE radius, $c_s$ is the isothermal sound speed, and $G$ is the gravitational constant \citep{Konyves2015}.  This critical mass does not take into account nonthermal turbulent motions; these can be considered by substituting the total velocity dispersion for the sound speed.  However, \citet{Andre2007} and \citet{Pokhrel2018} show that nonthermal motions are insignificant for low mass cores and produce unphysical formation efficiencies when treated as an effective pressure.  Therefore, we only consider the thermal BE mass.  Cores with a BE mass ratio $\alpha_{\rm BE} = M_{\rm BE, crit} / M_{\rm obs} \leq 2$ are considered self-gravitating and will eventually collapse and form protostars (generally these are cores above $M_{\rm BE, crit}$, while cores with $\alpha_{\rm BE} > 2$ do not have enough mass at that moment to remain bound.  
Though they may be pressure confined, they may eventually dissipate if more mass is not accreted.  

The separation between the median sizes of starred and starless cores, shown in Figure \ref{NEW_mass_size}, increases with age, changing from an average separation of $0.016 \pm 0.001$~pc to $0.044 \pm 0.002$~pc.  For RT2.2 and RT2.3, the majority of starred cores have sizes less than the median size of cores, with $60-80\%$ of small cores containing YSOs.  The number of starred cores less than the median overall size also increases as a function of age, though there is a slight turnover at 860 and 1000 pc for RT2.1.  Due to the low number counts of starred cores that lie within the center region at all distances in RT2.1, there is no statistically significant separation of starred and starless cores.  However, between RT2.2 and RT2.3, there is a $16 \pm 3\%$ increase in the number of starred cores with sizes smaller than the median.  

In order to derive synthetic core temperatures, we follow a similar approach that will be used by the C2C survey and is described in Section \ref{sec3}.  However, the large 36$\arcsec$ beam (ranging from a physical size of 0.11~pc at 650~pc away to 0.17~pc at 1000~pc away) will resolve different size scale temperature estimates depending on distance. 

If we assume the median temperature of the cores at each age and distance ($\sim15$~K), we find that the hotter cores are generally starred, gravitationally bound, and less extended compared to the starless cores which are large (and thus less dense), gravitationally unbound, and colder.
We also see a clear separation between the ratio of peak flux to total flux (the ``peakiness" of the core) of starred (diamonds) and starless (squares) cores.  Starred cores, which generally all fall on or above the BE line for all distances and ages, are smaller and less massive, but have higher S/N and are more concentrated (``peaky") compared to starless cores. As discussed in Section \ref{sec5}, this is because many of the older protostars are embedded in a massive disk and have little remaining envelope.  The fraction of starless cores above the BE line varies with evolutionary time and observational resolution, where early times show better agreement with the observed starless cores.

\subsection{Core Mass Function}
%% FIGURE 12
\begin{figure*}[tb]
\centering
\includegraphics[width=\linewidth]{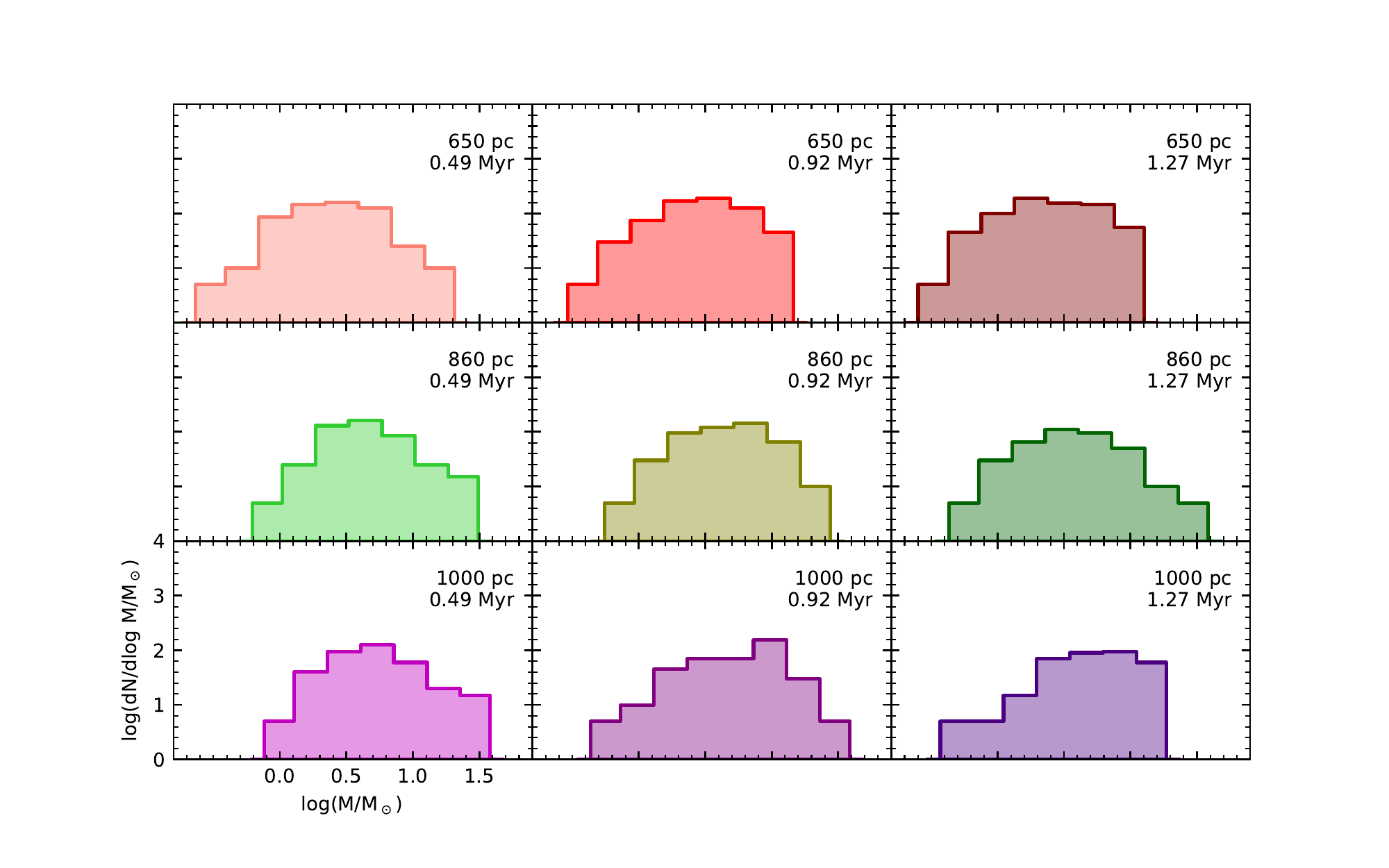}
 \caption{Core mass function for RT2.1 (left), RT2.2 (center), and RT2.3 (right) at 650 (top), 860 (middle), and 1000~pc (bottom) within the center 5~pc of each model.}
 \label{NEW_CMF}
\end{figure*}

We find that the average mass over time decreases for the starred cores, while remaining constant for the starless cores, resulting in CMFs with increased low mass bins.  This initial bias towards higher masses has been found in other molecular cloud simulations \citep{Smullen2020} and can be fit with a high mass slope $\mathrm{d}N/\mathrm{d}M \propto M^{-2}$ as seen by \citet{Guszejnov2015}. However, from 0.49~Myr to 0.92~Myr, the high mass slope decreases to $\mathrm{d}N/\mathrm{d}M \propto M^{-1.3}$ consistent with a Chabrier IMF, though shifted by a factor of $2.5-3.5$.  The number of starred cores also increases significantly; within the original 5~pc box, the starred cores increase from 2-3 to $\sim$30.  From 0.92~Myr to 1.27~Myr, the starred core counts remain fairly stable and spans a wide range of masses ($1-30$ M$_\odot$). However, we do not produce cores in the low mass regime ($<$~1~M$_\odot$), as radiative feedback in the simulation and lower resolutio of turbulence can inhibit small scale fragmentation \citep{Bate2012, Offner2009, Krumholz2011, Urban2010,Padoan2020}. The starless core tallies, while initially high due to a lack of sink particles in the simulation, remain fairly constant between the two later timesteps both in mass and number counts. 

Overall, the CMFs from $0.92 - 1.27$~Myr remain nearly constant between the two snapshots.  This consistency is seen in other simulations, such as \citet{Smullen2020} and \citet{Cunningham2018}. 
Relative consistency of the mass of the starless cores is also observed \citep[e.g.][]{Andre2014}. 
However, \citet{Smullen2020} also found significant variation over time in the properties of individual simulated isolated starless cores, even while the total CMF shape remained invariant.  Much of the core variation was due to the changing core definition, since the core identification method (dendrograms) was overly sensitive to small changes in the underlying core physical structure. This leads to concern that some algorithms, especially when used to identify cores in clustered regions, may not be robust. 

The reproduction of the combined starred and starless CMF, both in Figure \ref{pltsn08550_cmf_MS} and \ref{NEW_CMF}, is a good indication that the simulations are able to reproduce cores at similar masses and ages.  The consistency of the CMFs over this time range gives a good indication that the core identification and data reduction algorithms are able to identify the same cores with the same masses and sizes, confirming that systematic biases will not be a significant variable to untangle when analyzing the C2C core properties for most of the target clouds.  Only Perseus and Ophiuchus are closer then 400~pc in the C2C sample, and thus are likely to need special consideration for biases.

\subsection{Core Gas Correlation}

% FIGURE 13
\begin{figure*}[tb]
\centering
\includegraphics[width=\linewidth]{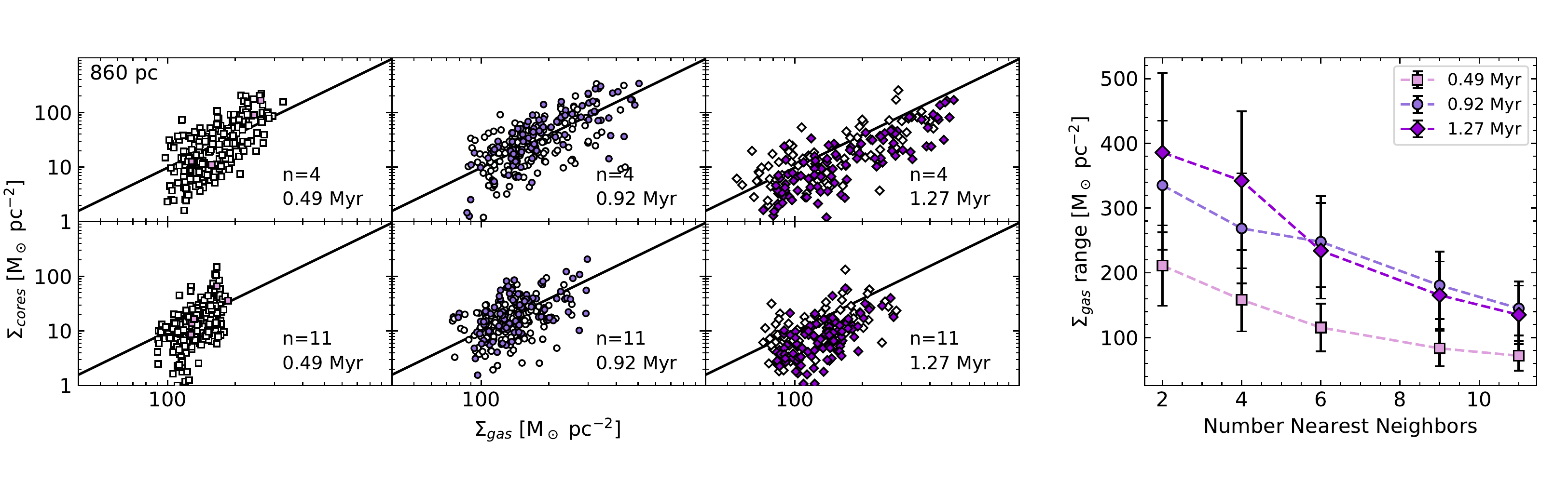}
 \caption{\textit{left} - Core gas correlation for RT2.1 (left/ squares), RT2.2 (center/circles), and RT2.3 (right/diamonds) at 860~pc for $n = 4$ (top/small scales) and $n = 11$ (bottom/large scales).  The black line corresponds to the best power-law fit from \citet{Sokol2019} with a power-law index of 1.99.  The white markers indicate starless cores, while the colored markers indicate starred cores. \textit{right} - The gas surface density range as a function of number of nearest neighbors ($n$) for each model shown left.  As $n$ increases, the gas surface density range decreases for all ages indicating clustering occurs only on the smallest scales.}
 \label{NEW_coregasrange}
\end{figure*}

%% FIGURE 14
\begin{figure}[tb]
\centering
\includegraphics[width=\linewidth]{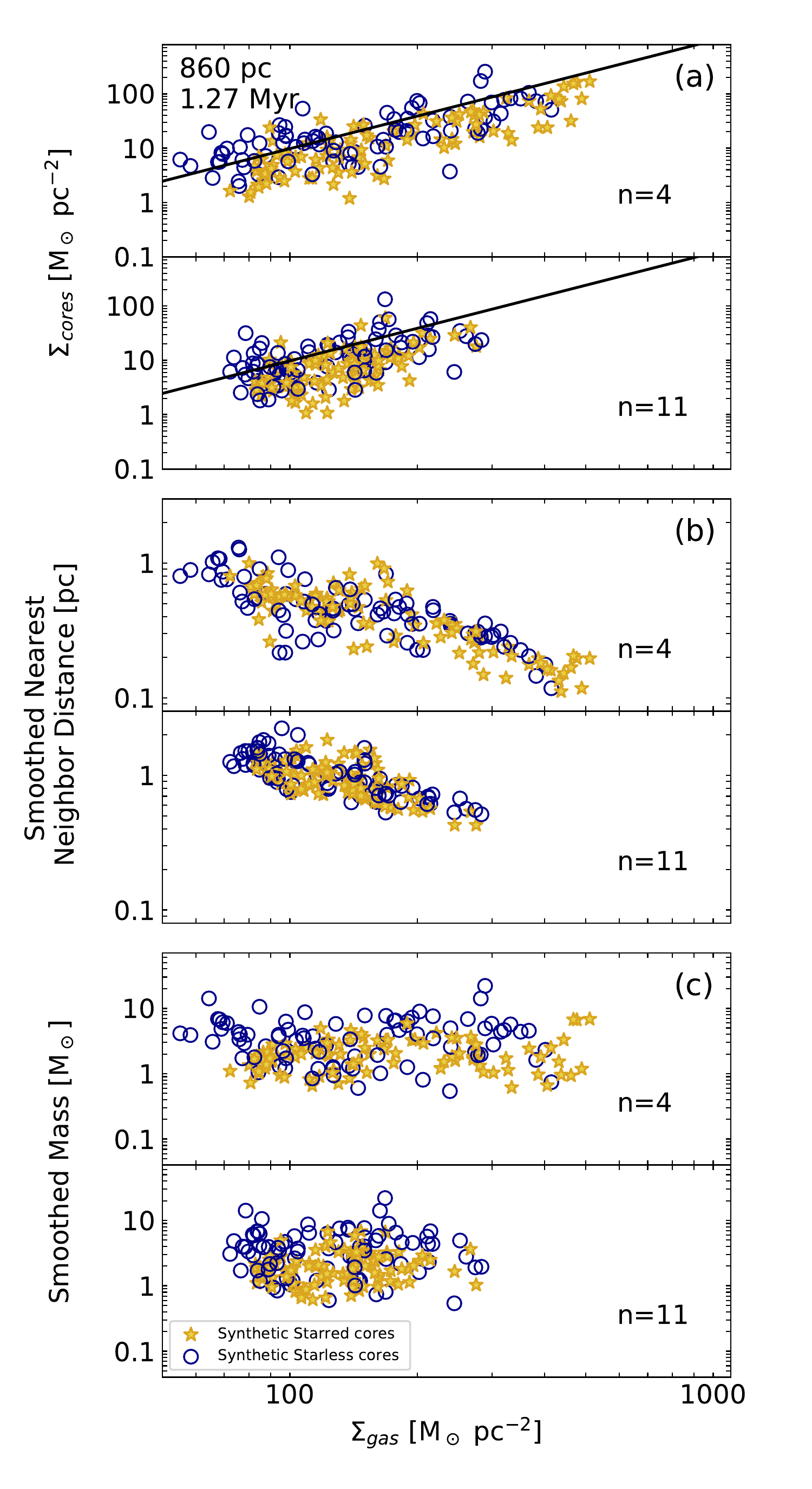}
 \caption{(a) Core gas correlation, (b) smoothing size scales, and c) smoothed mass for RT2.3 at 860~pc for $n=4$ (top panels/small scales) and $n=11$ (bottom panels/large scales).  The black line in a) corresponds to the bets power-law fit from \citet{Sokol2019} with a power-law index of 1.99.  The yellow stars are measurements centered on starred cores while the open blue circles are measurements centered on starless cores.  The starless cores are generally found to be more isolated, while starred cores are more clustered at large $n$.}
 \label{NEW_CGC_NND}
\end{figure}

We look at the core-gas correlation at the varying distances and ages to explore the effect of clustering and core formation over time within the simulations.  We find that the synthetic data overlap at all distances with very little variation in gas and core surface densities.  Therefore, in Figure~\ref{NEW_coregasrange} we show the core-gas correlation at 860~pc for all three ages as a representative sample.  The left panel of Figure~\ref{NEW_coregasrange} shows the core gas correlation at small ($n=4$) and larger scales ($n=11$) overlaid with the power-law fit (index of 1.99) from \citet{Sokol2019}, while the right panel shows the range of gas densities at various size scaling (number of nearest neighbors) at each age.  Overall, the synthetic data loci parallel and often overlap the power law fit at all ages and smoothing scales.  The range in gas densities decreases with increasing distance shifting towards low gas densities.  

The range of gas surface densities that host cores increases with age and shows higher gas column densities for the same core densities, indicating the gas has had time to accumulate due to the influence of gravity.  At younger ages, the gas is highly diffuse for all size scales indicating very few high gas density regions within the cloud.  Over time, the gas accumulates and collapses into small, high column density ``clumps" that form at most several cores, all while the gas remains relatively uniform at large scales.  This seeming inability to sustain larger molecular gas clumps may be a result of the driven turbulence and periodic boundary conditions of the simulation, which preclude largescale gravitational collapse and prevent large clumps from forming.

The simulated core-gas correlation falls along the same slope as found by \citet{Sokol2019} with a power law index of 1.99~$\pm$~0.03.  \citet{Sokol2019} found that this slope, which follows the model of thermal fragmentation \citep{Myers2009}, indicates the primordial gas distribution will be depleted quickly at high column densities due to the high mass efficiency of cores formed in that environment.  The extent that our synthetic cores extend in this core-gas density space is more limited than the observations, but the extent does grow toward higher core densities as the simulation evolves with time, especially noticeable in small-$n$ nearest neighbor measurements. When we separate the starred and starless cores (Figure~\ref{NEW_CGC_NND}), they are both well represented across the entire density range subtended, in agreement with the Mon~R2 observations. The main discrepancy appears to be the penchant for small number groupings in the simulations, resulting in the representation of the higher density regions being extremely sensitive to $n$ selection, seen in both the core density and nearest neighbor distance measurements (Figure~\ref{NEW_CGC_NND}a and b), unlike the observations of \citet{Sokol2019}.  Along with this discrepancy, we also find no variation in mass with gas density (Figure~\ref{NEW_CGC_NND}c), inconsistent with observations.  However, this disagreement is only seen at the low and high gas density ranges; the former a result of noise-induced false detections, and the latter due to temperature variations not taken into account in the observations.       

\subsection{Summary}
We have explored the robustness of synthetic observations with the same noise properties, filtering effects, and core identification algorithms as real observations in reproducing observed cores in order to assess the feasibility in disentangling environmental effects, age, and distance in core properties that will be observed by C2C.  We have found substantial over-filtering and over-segmentation at close cloud distances (D~$<$~300~pc and 500~pc, respectively) in synthetic observations based on AzTEC on the 32-m LMT.  The smaller TolTEC beam on the 50-m LMT may exacerbate the over-segmentation issue, while the larger field of view of the TolTEC arrays should reduce the impact of over-filtering at core scales when observing the nearest C2C target clouds.  However, for further distances where these effects are less significant, we are able to investigate observable evolutionary changes in core properties. We find that the synthetic observations are able to reproduce observed CMFs and produce cores with an efficiency that is consistent with the middle portion of the gas column density range of the Mon~R2 AzTEC survey's field of view.  However, clump formation in the simulation does not advance far enough in the simulation to reach the high gas column density and strong intermediate scale core clustering that is observed in Mon~R2.  Similarly, the simulations exhibit a growing separation in core masses and sizes in starred and starless cores with time, while no such distinction is observed in Mon~R2.  

These discrepancies between the observations and simulations limit the power of using synthetic observations to predict how observed core properties should behave when exposed to various environmental factors.   Next generation simulations such as STARFORGE \citep[e.g.,][]{grudic2020} are able to better capture physics from cloud to core size scales and thus may address some of the apparent discrepancies reported here. Similarly, future work on the observation side is needed to address the filtering and segmentation issues that plagued our analysis of nearby synthetic clouds.  An  observations simulator for TolTEC and LMT is under active development (Z. Ma, private communication) that includes end-to-end treatment of the telescope, optical and electronics system, and model atmosphere that will provide an  extremely useful test platform for future iterations of the C2C data treatment and cores analysis.

\section{Conclusions} \label{sec7}

In this paper, we began to explore the robustness of synthetic observations in accurately predicting molecular cloud core properties for the TolTEC Clouds to Cores Legacy Survey on the 50-m diameter LMT.  As this survey aims to map clouds at different ages and within different environments, assessing the feasibility of using synthetic observations to predict how environmental factors affect core properties and their evolution is essential.  

We produce synthetic 1.1~mm continuum emission observations by inserting snapshots of hydrodynamical radiative transfer simulations of star-forming regions into AzTEC/LMT-32m observations of the Mon~R2 cloud.  We use a python based image segmentation algorithm to find core candidates in synthetic observations and calculate various core properties, including mass, size, and temperature.  We explore a variety of simulation outputs at different ages and at varying distances in order to probe the full range of clouds that will be surveyed by C2C.  We find the following:

\begin{itemize}
\item Over-filtering and over-segmentation of cores occurs at distances less than 300~pc and 500~pc, respectively under the current data treatment and analysis path developed for C2C.  
The high resolution and better sampling that TolTEC will provide may improve or exacerbate these issues. Regardless, caution should be taken when analyzing cores from nearer distances because of these effects.  

\item Core masses, sizes, and the CMF found from a synthetic observation of simulation RT1 at 860~pc are consistent with the same observed core properties for cores in Mon~R2 found by \citet{Sokol2019}.  However, we find a separation in size between starred and starless cores that is not seen in the Mon~R2 observations, where the majority of synthetic cores with sizes $<$ 0.073 pc contain sink-particles/YSOs ($51 \pm 7\%$) compared to only $23 \pm 4\%$ of cores with sizes $>$ 0.073 pc. We expect this discrepancy is caused both by overly massive disks in the simulations, which appear as bright compact emission, and observational incompleteness at small core masses.

\item By sampling several ages of the RT2 simulations for synthetic observation, we explore %what evolution of 
the ranges of core mass, size, and clustering that are potentially observable.  We find starred cores decrease in mass and size, while starless cores remain invariant over time. In the simulations the synthetic starred cores move down the BE stability line with age as they accrete their envelope, separating from the starless cores.  

\item The simulated cores cluster where the gas is densest, but the groupings are universally small-$N$, exhibiting notable density reductions for modest increases in smoothing scale.  Their clustering character with respect to gas density is only consistent with the Mon~R2 observations within a narrow gas column density range.  The Mon~R2 observations exhibit core clustering correlated with gas structure at a wide range of gas column densities and smoothing scales.  These results suggest that while synthetic observations can reproduce CMFs, discrepancies between some simulations and observations of Mon R2 are substantial and require amelioration to improve confidence in predictions we derive from the simulations.  
\end{itemize}

\acknowledgments
RAG’s participation in this project was supported
by NASA ADAP grants NNX11AD14G, NNX15AF05G, and NNX17AF24G.  RAG and GW are supported in-part by the National Science Foundation, grant AST 1636621 in support of TolTEC, the next generation mm-wave camera for LMT. The AzTEC instrument was built and operated through support from NSF grant 0504852 to the Five College Radio Astronomy Observatory.  SSRO acknowledges funding support from NSF Career grant 1748571. This work is based on observations from the LMT, a project operated by the Instituto Nacional de Astrofísica, Óptica, y Electrónica (Mexico) and the University of Massachusetts at Amherst (USA), with financial support from the Consejo Nacional de Ciencia y Tecnología (project M0037-2018-297324) and the National Science Foundation (AST 2034318) respectively.  This work is based on observations made with Herschel, a European Space Agency cornerstone mission with science instruments provided by European-led Principal Investigator consortia and with significant participation by NASA.  This work is based [in part] on observations made with the Spitzer Space Telescope, which was operated by the Jet Propulsion Laboratory, California Institute of Technology under a contract with NASA.  This research made use of Photutils, an Astropy package for
detection and photometry of astronomical sources \citep{PHOTUTILS}.

\vspace{5mm}
\facilities{LMT(AzTEC), \textit{Herschel}(SPIRE)}

\software{astropy \citep{ASTROPY}, photutils \citep{PHOTUTILS}, Matplotlib \citep{MATPLOTLIB}, NumPy \citep{NUMPY}} 

\vspace{5mm}

\bibliography{reference}{}
\bibliographystyle{aasjournal}

\appendix

\section{Removing AzTEC Signal} \label{A1}
\begin{figure}[htb]
\plotone{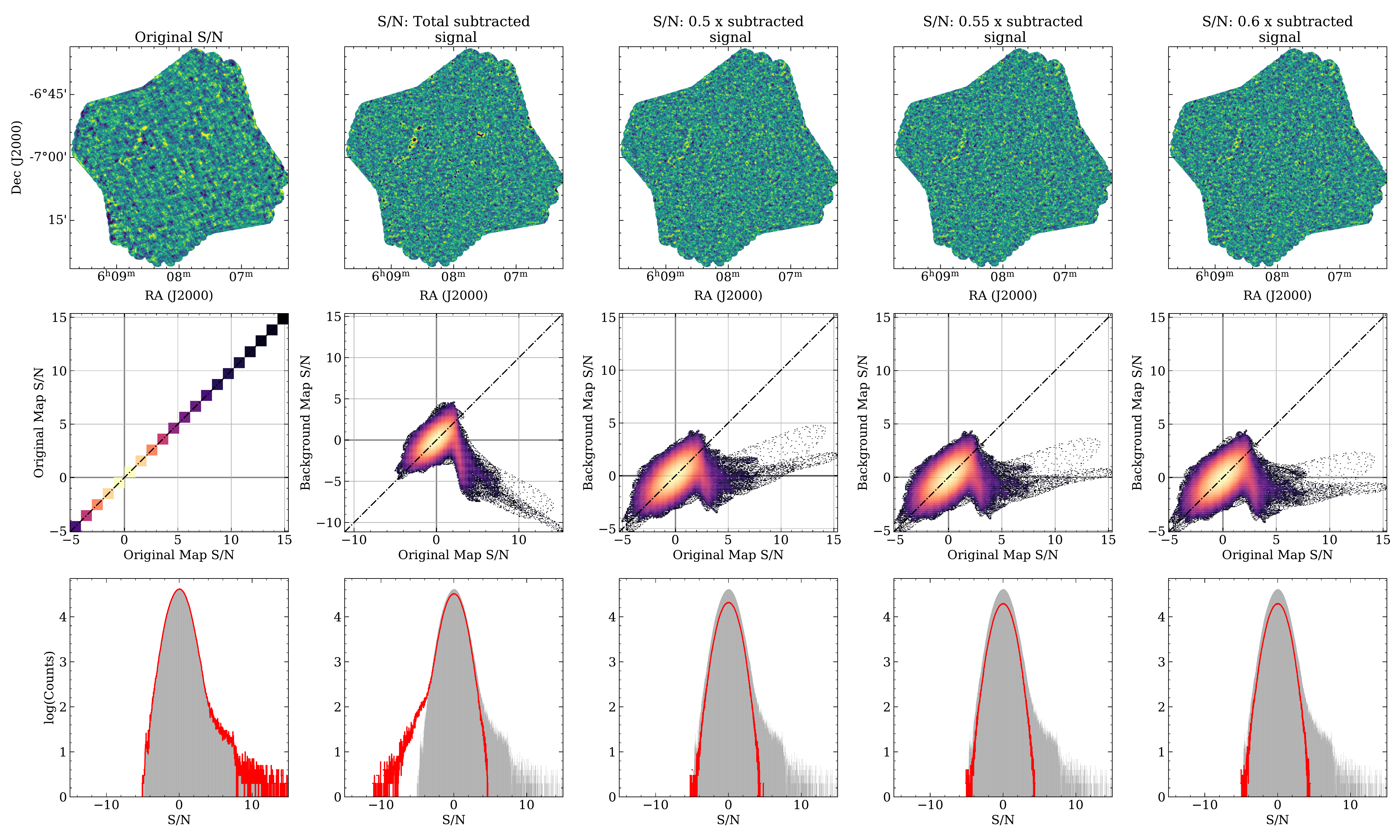}
 \caption{Effect of $f_{sub}$ on the negating signal in the Field~4 map.  \textit{top} - The resulting S/N background map for Field~4 with different factors of $f_{sub}$ (0, 1, 0.5, 0.55, 0.6) applied post \texttt{macana}. \textit{middle} - 2D histograms comparing the S/N of the original map to the background maps with various factors of $f_{sub}$ applied.  The dashed line is unity.  \textit{bottom} - Histograms of the S/N in the background maps (red) compared to the S/N of the original map (gray).  When significant signal is removed from the map, the background S/N is $-4 \lesssim \mathrm{S/N} \lesssim 4$.  When over-subtraction occurs, negative halos produce significant negative S/N values, and when under-subtraction occurs, positive signal produces significant positive S/N.  At $f_{sub} = 0.55$, the optimal amount of signal is subtracted out to diminish negative halos and strong positive signal.}
 \label{S2N_oversub_04}
\end{figure}

In order to insert the simulations into the AzTEC 1.1~mm continuum map, the original signal must be removed such that all the astronomical signal in the final synthetic observation is from the original simulation.  We find that a straight subtraction over-subtracts the original signal leaving negative halos surrounding the observed signal.  To address this, we determine the fraction of the original signal that should be subtracted (the over-subtraction fraction, $f_{sub}$) before inserting the simulation into AzTEC 1.1~mm continuum maps. We create negated AzTEC signal maps multiplied by various values of $f_{sub}$ that are added to the time stream and run through \texttt{macana}.  The goal is to create a map of well-behaved noise without strong correlated signal (or anti-signal).  Figure \ref{S2N_oversub_04} shows how the over-subtraction of signal in Field~4 changes for different $f_{sub}$ values.  When all the original signal is subtracted off, the background only map is significantly over-subtracted, resulting in negative S/N values for all original S/N $>$ 2.5.  However, when the original signal is halved ($f_{sub} \sim$ 0.5), the amount of over-subtraction is much less severe, with every pixel having a $|$S/N$| <$ 5. 

\begin{figure}[tb]
    \epsscale{0.65}
    \plotone{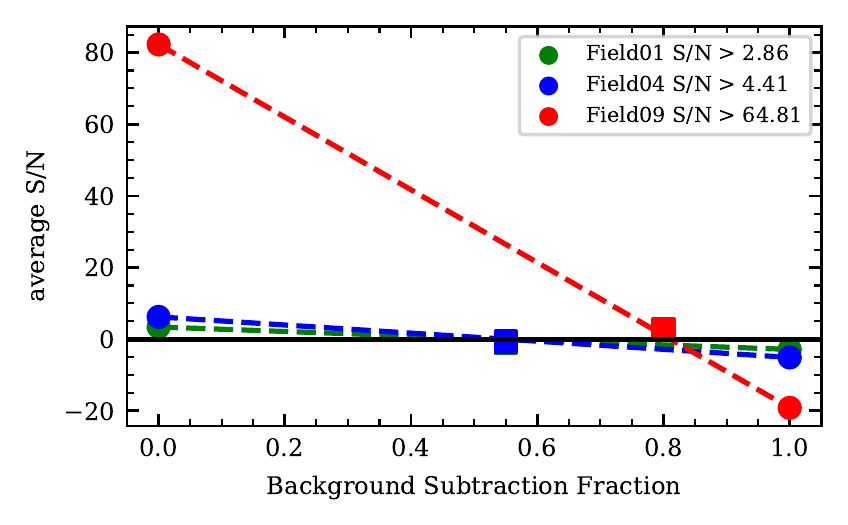}
    \caption{The effect of S/N on the oversubtraction fraction for three fields in Mon~R2: Field~1, Field~4, and Field~9.  Oversubraction increases with the increase of signal in the field; therefore, the oversubtraction fraction required to remove all the signal from the field also increases.}
    \label{S2N_oversub_149}
\end{figure}

To quantify $f_{sub}$, we find a linear relationship between $f_{sub}$ and the average S/N of pixels with S/N values greater than the pivot S/N.  The pivot S/N is found by:
\begin{enumerate}
\item Select all bins that have a background only S/N $\sim$ 0 
\item Determine the bins in that selection with counts greater than the median number of counts in the whole histogram. 
\item Find S/N$_{thresh}$: the original S/N value that corresponds to the rightmost bin with counts greater than the median of the whole histogram.  
\item Calculate the average S/N in each map with a S/N~$>$~N$_{thresh}$.
\end{enumerate}
This average S/N is the S/N of all pixels in which the original S/N was large enough to be affected by the over-subtraction. 
With this relationship, we can then find $f_{sub}$ where the average S/N of those affected pixels will be zero and create a blank canvas in which to add the synthetic emission map.  As shown in Figure \ref{S2N_oversub_149}, for the three fields we tested (Field 01 (relatively empty), Field~4, and Field 09 (relatively strong emission)), we found a range of $f_{sub}$ from 0.55 (Field 01) - 0.8 (Field 09), indicating fields with more emission have a higher $f_{sub}$.  

\section{Image Segmentation}  \label{ImSeg_appendix}
At submillimeter wavelengths, the thermal emission from cores is optically thin, allowing the full flux density to be measured.  However, how these cores are identified is still an issue that must be resolved before core properties can be measured.  Beyond searching by eye, several clump finding algorithms exist that search for bright peaks \citep[ClumpFind;][]{Williams1994}, ascend gradients until reaching a peak \citep[FellWalker;][]{Berry2015}, or use hierarchical branches to find a range of clumps \citep[Dendrograms;][]{Rosolowsky2008}.  Each of these methods have biases and work in different ways.  However, none take into account the noise properties and maps produced from the data reduction pipeline from AzTEC and eventually TolTEC.

\texttt{macana}, the AzTEC C++ Data Reduction Pipeline, produces a co-added signal, weight, S/N and psf maps.  To characterize the noise in each scan of the time-stream data, noise realization maps are produced through a jackknifing technique.  These produce a good estimation of the noise variance in each scan and can give a false detection rate when finding clumps.  In order to utilize the noise realization maps to the fullest extent and get an accurate core census, \citet{Sokol2019} used a core identification algorithm similar to ClumpFind for both the signal map and the noise realization maps in order to identify real and false cores.  This algorithm was written in IDL and used the IDL watershed algorithm to search for continuous emission.  
With the upcoming TolTEC Clouds to Cores Survey (C2C), this algorithm has been updated, converted into a working python based library (named ImSeg\footnote{http://github.com/sbetti22/ImSeg}), and documented so that it can be used by the C2C community.      
 
ImSeg is a core detection and characterization utility that uses the photutils package 
\texttt{photutils.Segmentation} to detect and deblend clumps and cores in both observational data and simulations.
After detecting core candidates, ImSeg measures the characteristics of each core, including RA/Dec, total and peak flux, area, FWHM semi-major/minor axes, and mass.  All core candidates are then put through a ``goodness" test using the noise realization maps to determine the final core catalogue and CMF.

ImSeg is currently set up for 350, 450, 850, and 1100~$\mu$m wavelength observations and has been successfully applied to SCUBA-2 450 and 850~$\mu$m data as well as LMT AzTEC 1100~$\mu$m data.

Before identifying the cores, the flux map is first masked on values with a S/N $<$ 2.5 and/or where the weight was less than 40 percent of the median non-zero weight values of the map.  This masking allows only significant signal to be identified.  A noise map is also created from the signal and S/N map in order to give a threshold level for detection.  The masked signal map is then run through \texttt{photutils.Segmentation detect\_sources()} algorithm which detects blobs with values greater than the noise at each pixel.  For a blob to be identified, it must have a minimum number of 8-connected pixels (npixels; pixels touching along edges or corners).  This parameter is tuneable; however, we choose a value of npixels which corresponds to blobs slightly smaller than 0.05~pc.  As 0.05~pc is the average core size, by choosing an npixel slightly smaller, we include cores while excluding potential noise peaks or hot pixels.  However, as this method only determines connected pixels, overlapping blobs will be identified as one component.  Therefore, we utilize the \texttt{photutils.Segmentation deblend\_sources()} multi-thresholding and watershed algorithm to separate individual cores.  This method requires two parameters: nlevels and contrast.  nlevels refers to the number of multi-thresholding levels, while contrast is the fraction of the total flux that a peak must have to be its own blob.  The number of levels is used for every source; this means that 2$\sigma$ sources are separated by the same number of levels as a $10\sigma$ source.  This poses a problem as either over-segmenting can occur (nlevels is too high) or undersegmenting (nlevels is too low).  To get around this issue, we make several changes to the \texttt{deblend\_sources()} function.  Before running the function, we create a noise-based contour image of our signal map with steps of 1$\sigma$ in size.  The contour image is found by scaling the masked signal map by the median noise value in the masked pixels and then the fractional values are truncated to yield the noise-based contour image. This image is stored as a greyscale byte image.

The \texttt{deblend\_sources()} function is changed to use this noise based contour map to find the number of multi-thresholding levels to use for each and their corresponding data values.  Each blob is then segmented with a different number of levels based on the contour image; however, the step size between thresholding levels for each blob is always 1$\sigma$.  

After each core is deblended, the properties of each core are determined.  For each core, \texttt{photutils.Segmentation source\_properties} calculates several properties.  We utilize the central RA and Dec, peak S/N and signal, and area in pixels squared, and flux values within each core.  With these various properties, we find and calculate: the central RA and Dec, peak S/N, peak signal, total S/N, total signal, area in degrees squared, the half peak power area, half peak power central RA and Dec, and determine if the core is too close to the edge of the map. Additionally, we calculate the temperature of each core, and the number of YSOs within the footprint of the core.

The final core candidates are selected using a ``goodness" test based on the whether the core passes three tests: minimum S/N threshold, ``good" score minimum threshold, and a minimum column density ratio threshold.  

The column density from the emission map is calculated as:
\begin{equation}
N(H_2) = \frac{S_\nu}{B_\nu(T) \kappa_\nu \mu_{H_2} m_H \Omega_B}
\end{equation}
where S$_\nu$ is the total signal from each core in Jy, $B_\nu(T)$ is the Planck function at the temperature of each core in K, $\kappa_\nu$ is the dust opacity, $\mu_{H_2}$ is the mean molecular weight, which we take to be 2.8 \citep{Kauffmann2008}, and $\Omega_B$ is the beam area.  We assume a constant dust opacity interpolated for various wavelengths taken from model 5 of \citet{Ossenkopf1994} for thin ice mantles.  For 350, 450, 850, 1100 $\mu$m, we assume 0.101, 0.0674, 0.0114, 0.0121 cm$^{2}$ g$^{-1}$, respectively.  We also assume a gas to dust ratio of 100.

As shown by \citet{Sokol2019}, false cores from noise realization maps are poor matches with the diffuse gas from an inputted column density map, while high S/N core candidates correlate well (their column density ratio is approximately unity).  Therefore, by finding cores that occupy the same S/N - column density ratio parameter space as the false noise-realization cores, we can separate the false detections from cores.  

To do this, we create a 2D histogram of the S/N and column density ratio for both the noise realization false cores and the core candidates.  The ratio of these two histograms is found in order to determine the parameter space where false detections are likely to occur.  Confidence intervals are calculated for the ratio histogram. If a core candidate lies within the ``good" score minimum confidence interval, then it is considered a false detection.  Cores that lie outside this confidence interval have a low probability of being considered a false detection.  The final census of core candidates is selected from cores that lie outside this ``good" score confidence interval, have a S/N above the minimum threshold, and have a column density ratio greater than the provided threshold.  If no noise realizations are provided, only the final two criteria are used to create the final catalogue.   

\subsection{Comparison to Sokol et al. (2019) IDL implementation}

ImSeg is based off of a similar implementation outlined in \citet{Sokol2019} written in IDL using IDL watershed.  In order to gauge the effectiveness and reliability of ImSeg, we compare the two methods in order to see if we can produce the same results. 

\begin{figure}[h]
    \centering
    \begin{minipage}{\textwidth}
        \epsscale{0.7}
        \plotone{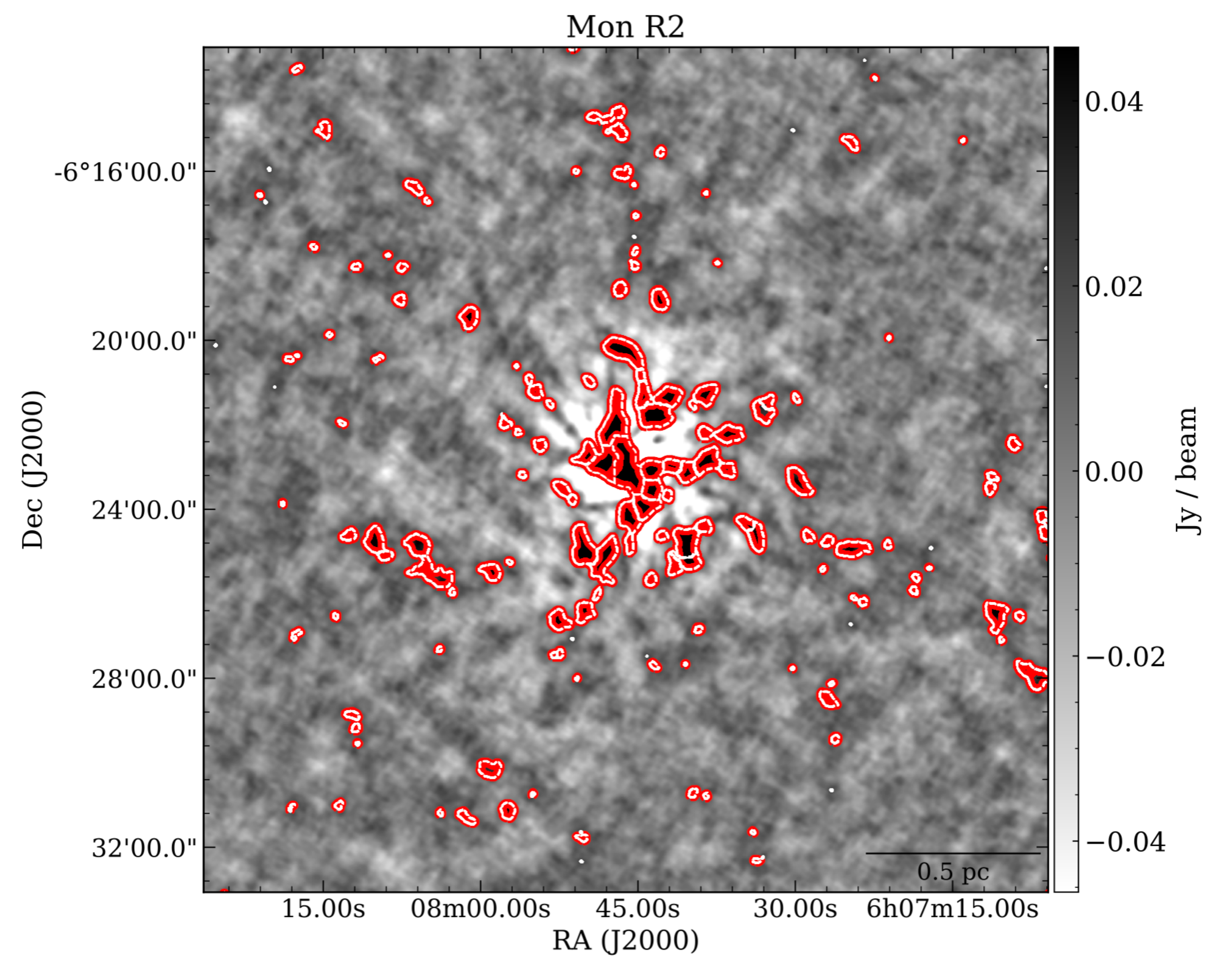}
    \end{minipage}
    \begin{minipage}{\textwidth}
        \plottwo{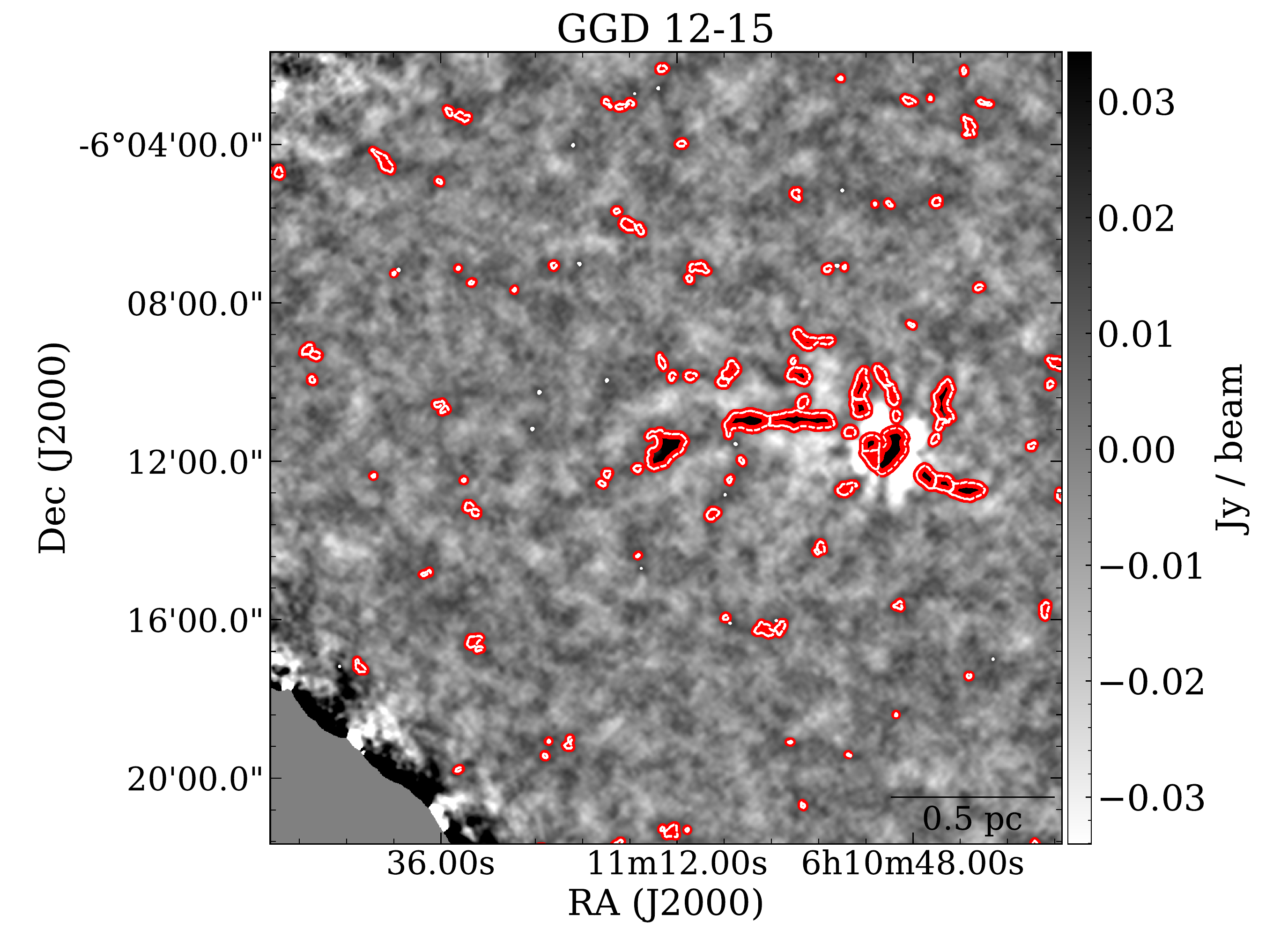}{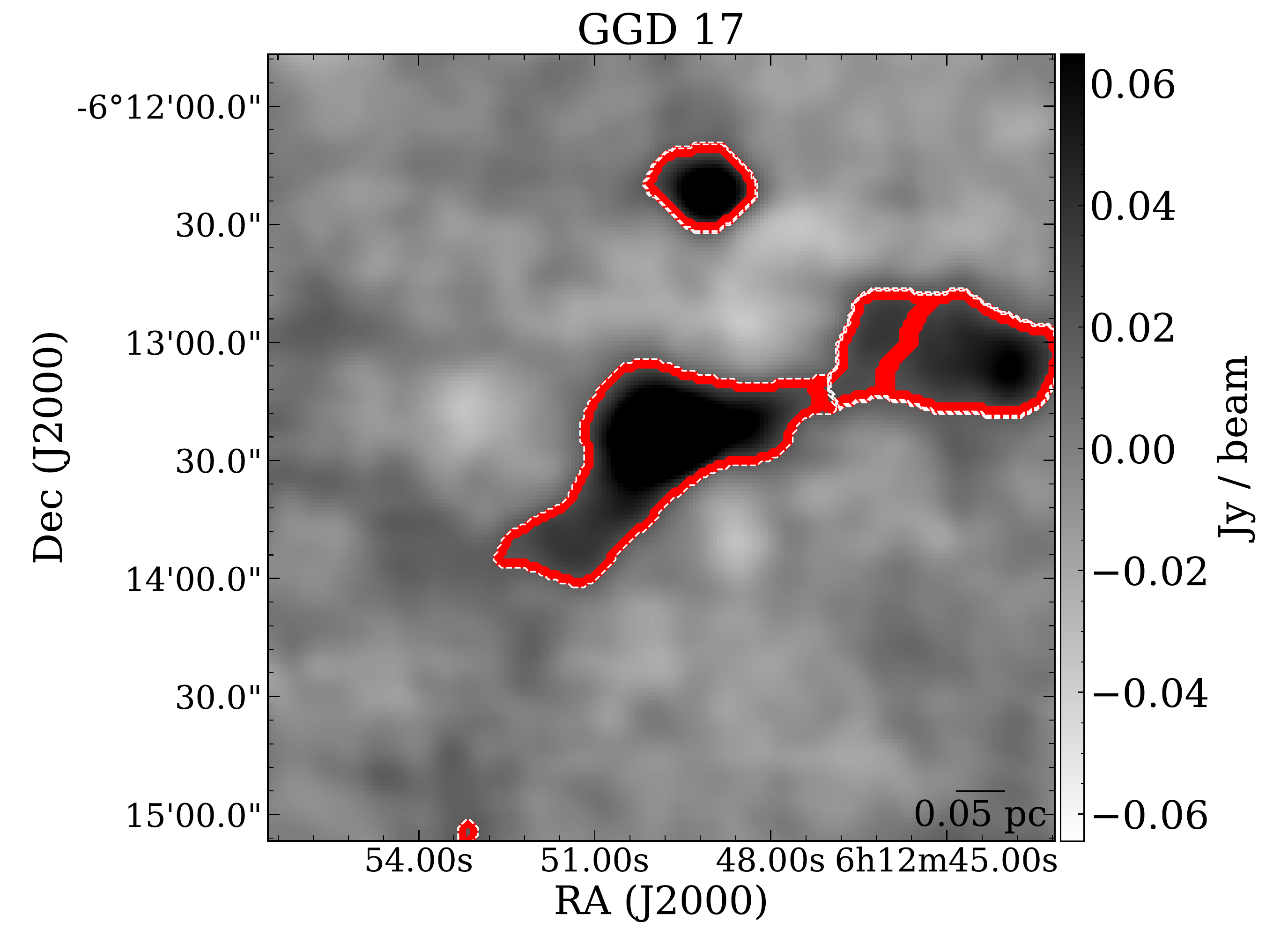}
    \end{minipage}
    \caption{Footprint matching between ImSeg (red) and IDL watershed \citep[white;][]{Sokol2019} on top of 1.1 mm AzTEC emission map for the Mon~R2 region (top), GGD 12 15 (bottom left) and GGD 17 (bottom right) region.  These main areas of the molecular cloud are shown in order to give a range of core densities.}
    \label{footprint_matching}
\end{figure}

We use 1.1 mm data Mon~R2 giant molecular cloud data taken with AzTEC on the Large Millimeter Telescope Alfonso Serrano (LMT) from 2014 November 27 - 2015 January 31.  Fourteen fields were observed covering an area of 2 deg$^2$ and were reduced with \texttt{macana}. We then applied ImSeg to the reduced maps in order to identify the cores within the cloud.  We used \textit{Herschel} 500 $\mu$m derived column density and temperature maps from \citet{Pokhrel2016}.   These cores were then compared to the original core catalogue and footprints found in \citet{Sokol2019} using IDL.  In Figure \ref{footprint_matching}, we show a comparison between the IDL method and ImSeg for the main clusters within the Mon~R2.  Red is the footprint found with ImSeg while white is the footprint found with IDL watershed.  By visually inspecting both the dense cluster and more isolated cores, we see that ImSeg reproduces the same footprints as IDL watershed. 
%% FOOTPRINT IMAGES

\begin{figure}[tb]
    \centering
    \plotone{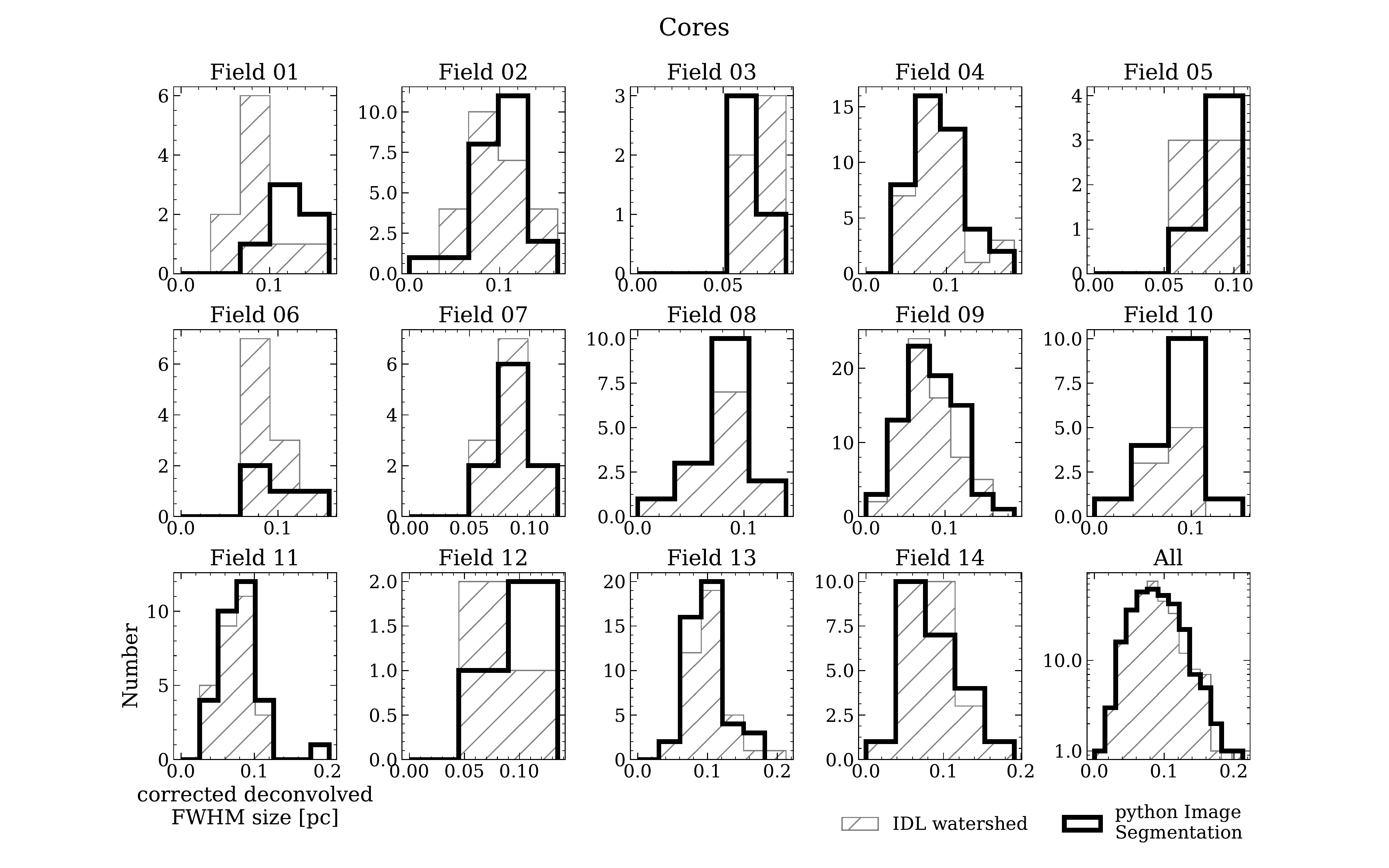}
    \caption{Histogram of the corrected deconvolved FWHM size for all cores within each field and all combined found with ImSeg (black) and IDL watershed \citep[gray hatch;][]{Sokol2019}.}
    \label{size_hist}
\end{figure}

We then look at the core properties for all fields, specifically area, size, and mass, to see how the two methods compare.  Figure \ref{size_hist} shows the deconvolved FWHM size of all cores found with ImSeg that passed the threshold cuts.  Overall, we found 306 cores with ImSeg and found $246 \pm 9$ matches ($83 \pm 3 \%$) in the original 295 cores IDL catalogue.  Therefore, there were $\sim 50$ IDL cores not found with ImSeg, and $\sim 60$ ImSeg cores not found with IDL.  The majority of non-matches occurred for cores with predominantly small or large areas.  This is due to either ImSeg or IDL watershed not breaking up a core candidate into multiple candidates, giving candidates with overall larger areas, or small area candidates not passing either the watershed algorithm or the threshold cuts.     

\begin{figure}[tb]
    \centering
    \epsscale{0.8}
    \plotone{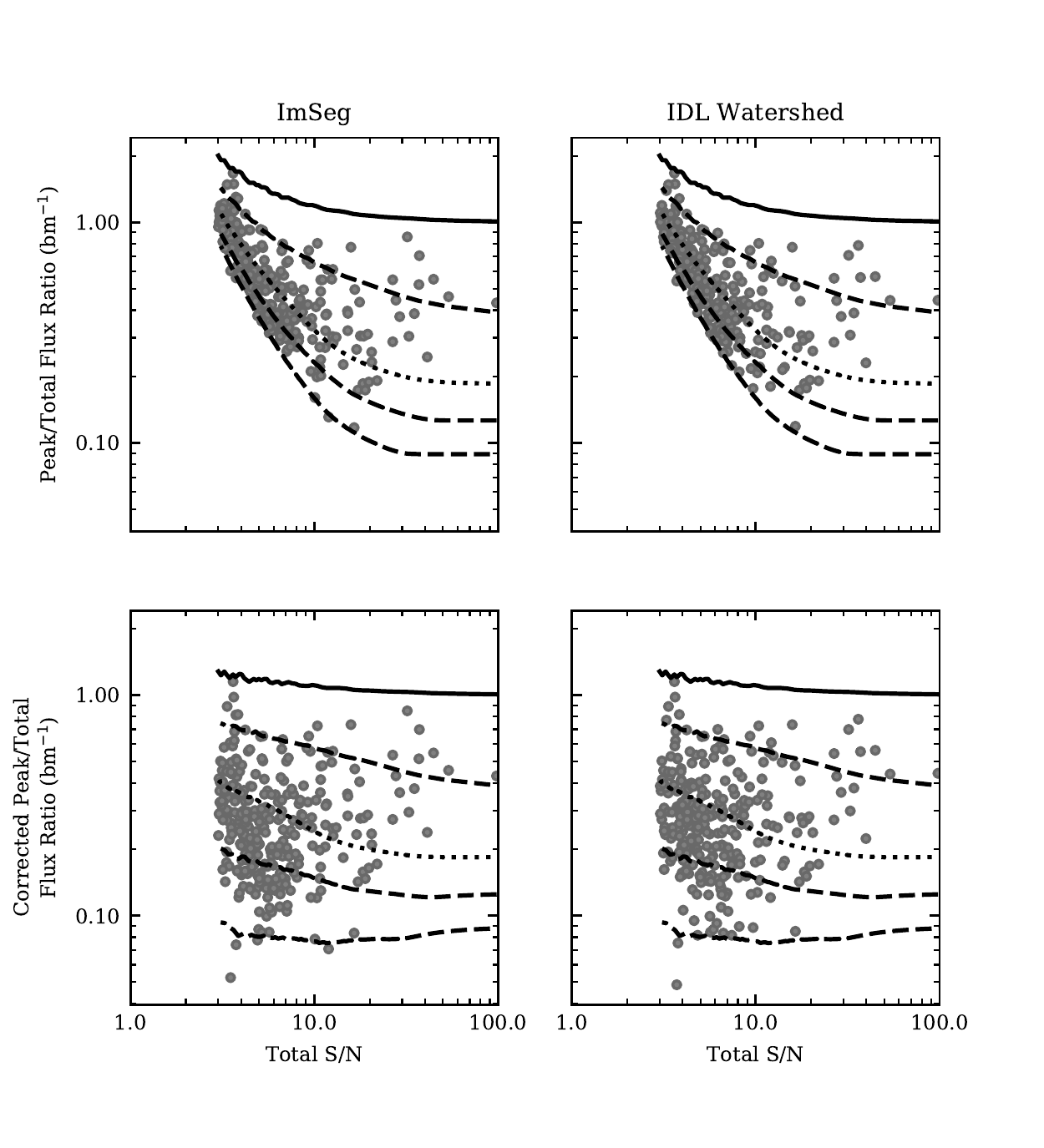}
    \caption{Peak-to-Total flux vs total S/N for cores found with ImSeg (left) and IDL watershed \citep[right;][]{Sokol2019}.  The top panels show the uncorrected flux ratio while the bottom panels show the flux ratio corrected for the noise bias.  The black dashed lines are the three Plummer-like models for with a power-law index of 2 and scale lengths of 2, 12, and 18 arcseconds (top to bottom), the black solid line is the final beam profile, and the black dotted line is composite-core profile.}
    \label{fluxratioSN}
\end{figure}

Finally, we reproduce diagnostics from \citet{Sokol2019} in order to make sure the core properties are similar.  As shown in Figures 7-8 of \citet{Sokol2019}, the 1.1 mm AzTEC Mon~R2 survey is shallow so cores are spatially resolved and lower S/N cores are not fully detected.  \citet{Sokol2019} corrected the underestimation of total flux and core mass by modeling and characterizing the peak to total flux ratio relation and then correcting for high ratios at low S/N.  They tested their correction by constructing several Plummer-like models that range the peak-to-total flux and total S/N parameter space, as they fit the radial profile of pre-stellar cores.  We applied these same models and corrections to confirm that ImSeg gives the same noise bias and that the same corrections can be used to correct the underestimation of the flux (Figure \ref{fluxratioSN}).  

\begin{figure}
    \centering
    %% Mass vs Size
    \begin{minipage}{\textwidth}
    \includegraphics[width=\textwidth]{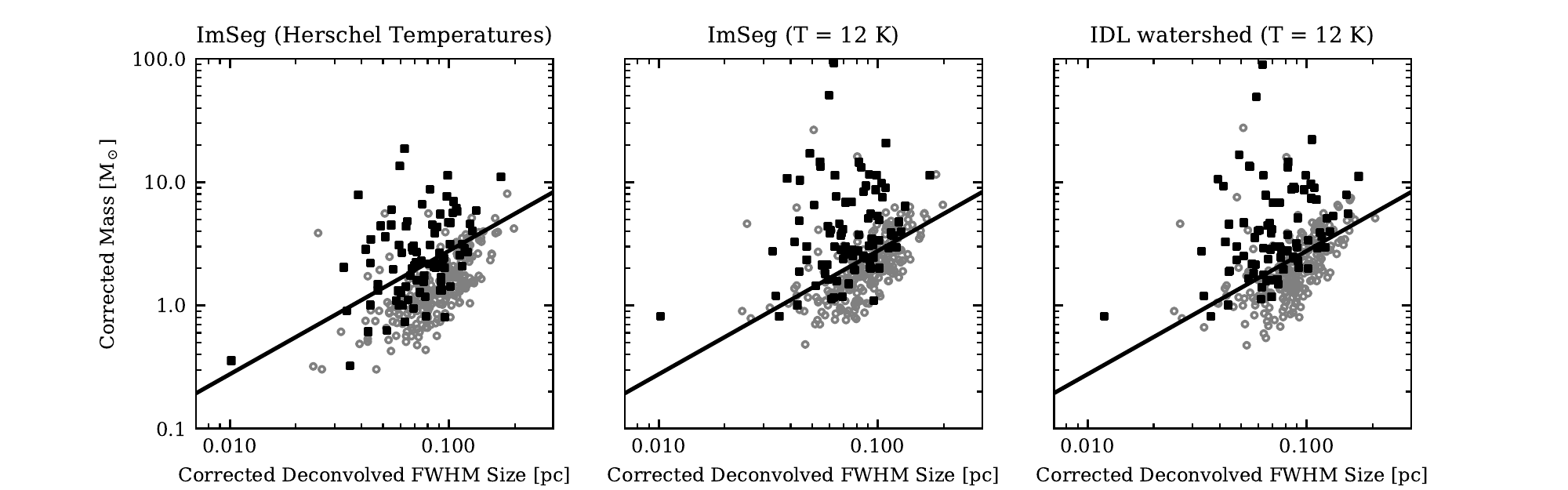}
    \caption{Mass vs deconvolved FWHM size relation for core found with ImSeg (left and center) and IDL watershed \citep[right;][]{Sokol2019}.  The left panel shows the cores found with ImSeg where mass was calculated using the average temperature of the core found from \textit{Herschel}.  The center panel shows the cores found with ImSeg where mass is calculated assuming 12 K, the temperature assumed in \citet{Sokol2019}.  The right panel shows the cores found with IDL watershed from \citet{Sokol2019}.  The black line in all panels is the Bonner-Ebert line for cores with T = 12 K.  The black squares represent cores starred cores, while the gray circles are starless cores.}
    \label{mass_size}
    \end{minipage}%
    %% CMF
    \begin{minipage}{\textwidth}
    \includegraphics[width=\textwidth]{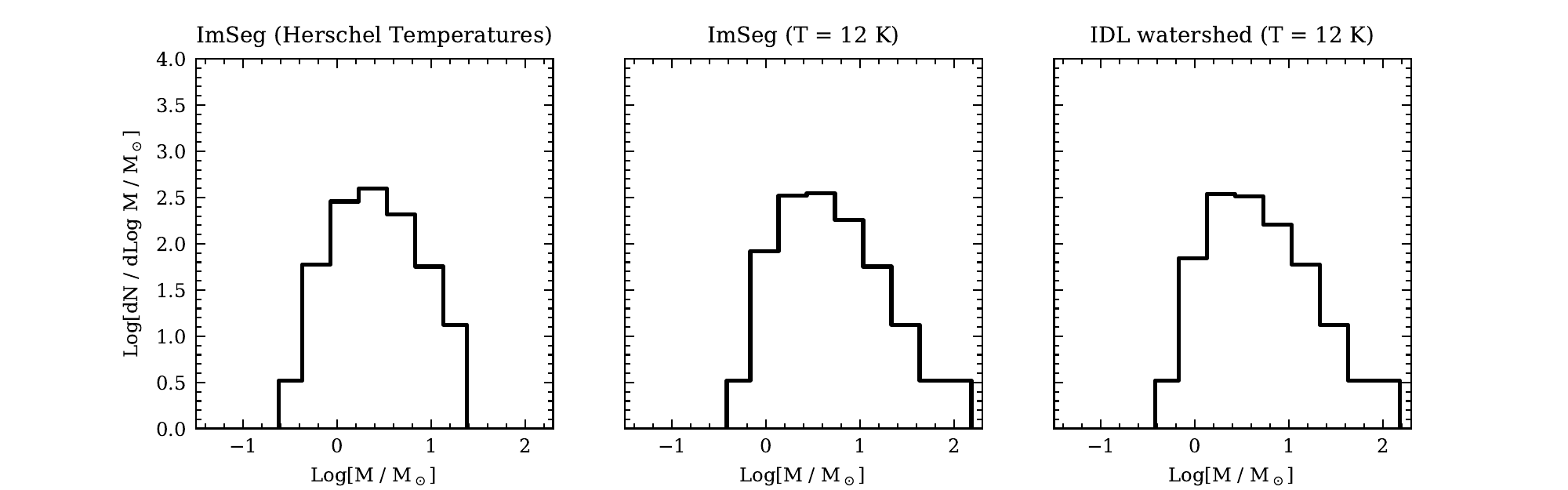}
    \caption{Mass corrected core mass function found with ImSeg (left and center) and IDL watershed \citep[right;][]{Sokol2019}.  The panels are as in Figure \ref{mass_size}.  Masses less than 3 M$_\odot$ have incomplete sampling.}
    \label{CMF}
    \end{minipage}%
\end{figure}

Then, we applied the flux correction to correct the core mass in order to look at the mass vs size relation (Figure \ref{mass_size}) and the core mass function (Figure \ref{CMF}).  As we used a \textit{Herschel} temperature map to derive the temperatures of the cores, the masses will be different from the masses found from \citet{Sokol2019} who assumed T = 12 K for all cores.  We therefore recalculate the masses assuming this temperature to confirm the reproducibility of these diagnostics.  Figures \ref{fluxratioSN} - \ref{CMF} confirm visually that ImSeg reproduces the same properties as IDL watershed.

% % %% SIZE HISTOGRAM

%% FLUXRATIO VS SN

\section{Disks within synthetic starred cores} \label{App3}

The starred synthetic cores are generally less massive and smaller than starless cores.  This is due to the addition of bright, massive disks within the simulation.  When these are smoothed and run through the \texttt{macana} pipeline, they appear as ``peaky" small cores encompassing a sink particle.  These massive older disks are a common occurrence in HD simulations \citep[e.g.][]{Zhao2020} and help account for the discrepancy between observations and the synthetic cores described in Section \ref{sec5}.          

\begin{figure}[tb]
    \centering
    \epsscale{0.6}
    \plotone{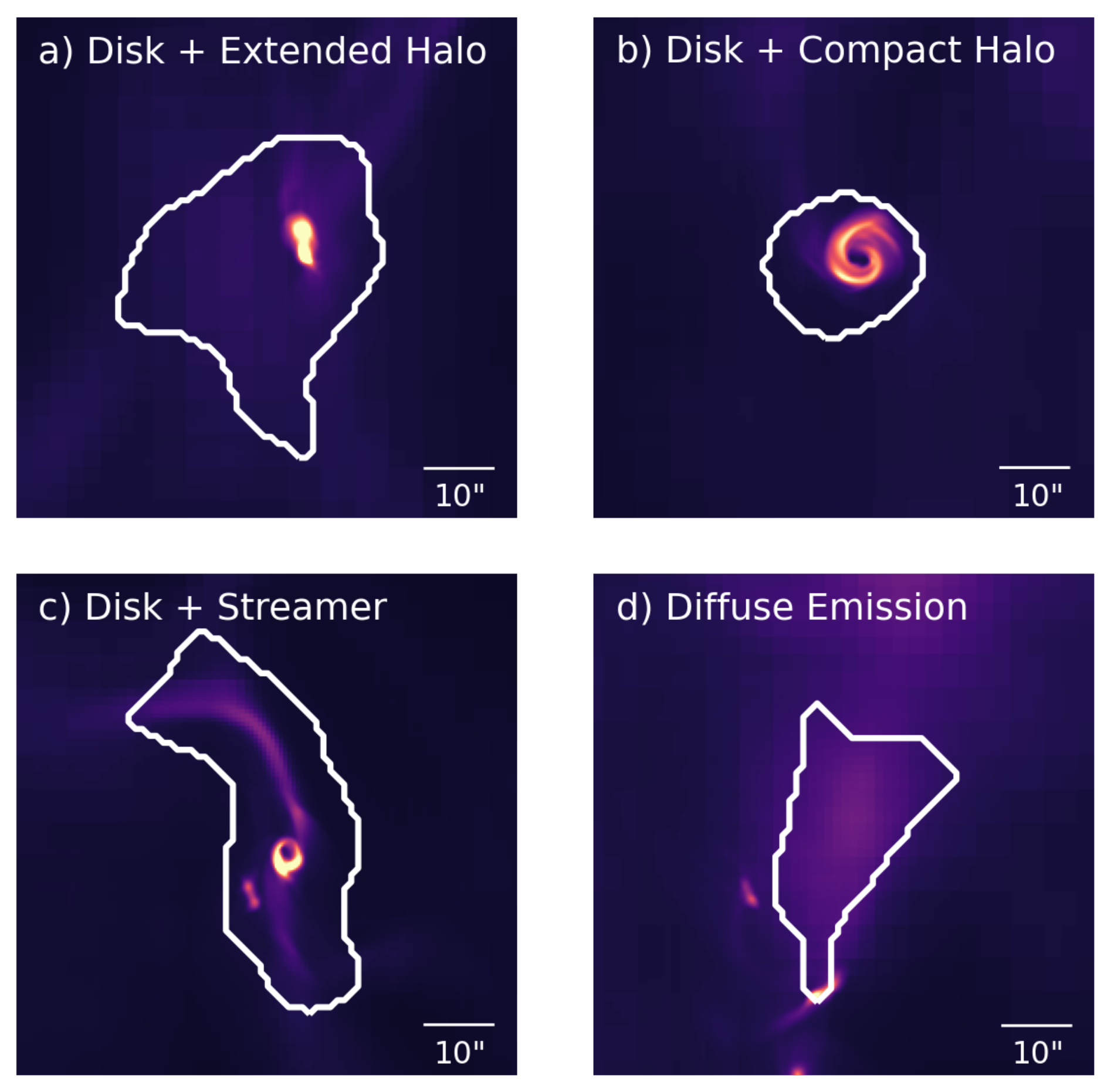}
    \caption{Representative sample of the different types of simulated starred cores overlaid by their synthetic observed core boundary (white contour): a) disk and extended halo, b) disk with compact halo, c) disk with streamers, and d) diffuse emission.   }
    \label{postage}
\end{figure}

By inspecting the original simulation, we find four main types of starred cores: a) those with a disk surrounded by an extended halo, b) those with a disk surrounded by a compact halo, c) a disk with streamers, and d) only diffuse emission where the disk is not visible (but a sink particle is still located within the core boundary).  We illustrate these four types by a representative sample in Figure \ref{postage}.  

\end{document}